\documentclass[journal,twoside]{IEEEtran}
\usepackage{cite}
\usepackage{graphicx}
\usepackage[cmex10]{amsmath}\interdisplaylinepenalty=2500
\usepackage{relsize}
\usepackage{hhline}
\usepackage{nicefrac}
\usepackage{amssymb,bm,upgreek}
\usepackage[table]{xcolor}
\usepackage{multirow}
\usepackage[labelformat=simple]{subcaption}
\usepackage{caption}
\usepackage{xcolor}
\newtheorem{theorem}{Theorem}

\newtheorem{proposition}{Proposition}

\begin{document}
\title{Max-Min Power Control in Downlink Massive MIMO with Distributed Antenna Arrays}

\author{Noman~Akbar,~\IEEEmembership{Member,~IEEE,}
        Emil Bj\"ornson,~\IEEEmembership{Member,~IEEE,} \\ Nan~Yang,~\IEEEmembership{Senior Member,~IEEE,} and ~Erik G. Larsson, \IEEEmembership{Fellow,~IEEE}%
\thanks{N.~Akbar and N.~Yang are with the Research School of Engineering, Australian National University, Canberra, ACT 2600, Australia (e-mail: \{noman.akbar, nan.yang\}@anu.edu.au). E.~Bj\"ornson and E.~G.~Larsson are with the Department of Electrical Engineering (ISY), Link\"oping University, Sweden (e-mail:\{emil.bjornson, erik.g.larsson\}@liu.se)}
\thanks{This research work was performed when the first author visited the Division of Communication Systems, Department of Electrical Engineering (ISY) in Link\"oping University, Sweden.}
\thanks{This work was supported by the Australian Government Research Training Program (RTP) Scholarship, the Swedish Research Council (VR), and ELLIIT. The work of N.~Yang was supported by the Australian Research Council's Discovery Project (DP180104062).}
\thanks{This work was presented in part at the IEEE International Conference on Communications (ICC), Kansas City, MO, May 2018, which appears in this manuscript as reference [22].}}

\markboth{IEEE TRANSACTIONS ON COMMUNICATIONS}{Akbar \MakeLowercase{\textit{et al.}}: Max-Min Power Control in Downlink Massive MIMO with Distributed Antenna Arrays}

\maketitle

\begin{abstract}
In this paper, we investigate optimal downlink power allocation in massive multiple-input multiple-output (MIMO) networks with distributed antenna arrays (DAAs) under correlated and uncorrelated channel fading. In DAA massive MIMO, a base station (BS) consists of multiple antenna sub-arrays. Notably, the antenna sub-arrays are deployed in arbitrary locations within a DAA massive MIMO cell. Consequently, the distance-dependent large-scale propagation coefficients are different from a user to these different antenna sub-arrays, which makes power control a challenging problem. We assume that the network operates in time-division duplex mode, where each BS obtains the channel estimates via uplink pilots. Based on the channel estimates, the BSs perform maximum-ratio transmission in the downlink. We then derive a closed-form signal-to-interference-plus-noise ratio (SINR) expression, where the channels are subject to correlated fading. Based on the SINR expression, we propose a network-wide max-min power control algorithm to ensure that each user in the network receives a uniform quality of service. Numerical results demonstrate the performance advantages offered by DAA massive MIMO. For some specific scenarios, DAA massive MIMO can improve the average per-user throughput up to $55\%$. Furthermore, we demonstrate that channel fading covariance is an important factor in determining the performance of DAA massive MIMO.
\end{abstract}


\section{Introduction}
Massive multiple-input multiple-output (MIMO) is a key technology in the fifth generation (5G) wireless standard \cite{Parkvall17}. In massive MIMO networks, the base stations (BSs) are equipped with a very large number of antennas \cite{Marzetta2010,Larsson2014,Yang2015,Akbar2016a,Marzetta2016}. Massive MIMO offers high spectral efficiency, energy efficiency, and throughput \cite{Nguyen2017,Akbar2018a}. However, practical aspects related to the deployment of massive MIMO in 5G networks are still relatively unexplored. Most existing research in massive MIMO has treated deployments with all base station antennas co-located in a compact antenna array \cite{Marzetta2016,Liu2014,Emil2016b,Akbar2016a,Akbar16b}.

In a co-located deployment, the path-loss between a user and all the antenna elements within a cell is assumed to be the same. This is
because of the fact that, the base station antennas in the array are assumed to be relatively close to one another and have the same radiation patterns. For this deployment scenario, performance has been characterized analytically for independent Rayleigh fading \cite{Marzetta2016} and for correlated Rayleigh
fading \cite{Emil2017bb}, and efficient power control algorithms are available.

\begin{figure}[!t]
\centering
\begin{subfigure}[b]{0.5\textwidth}
\centering
{\includegraphics[width=3.4in]{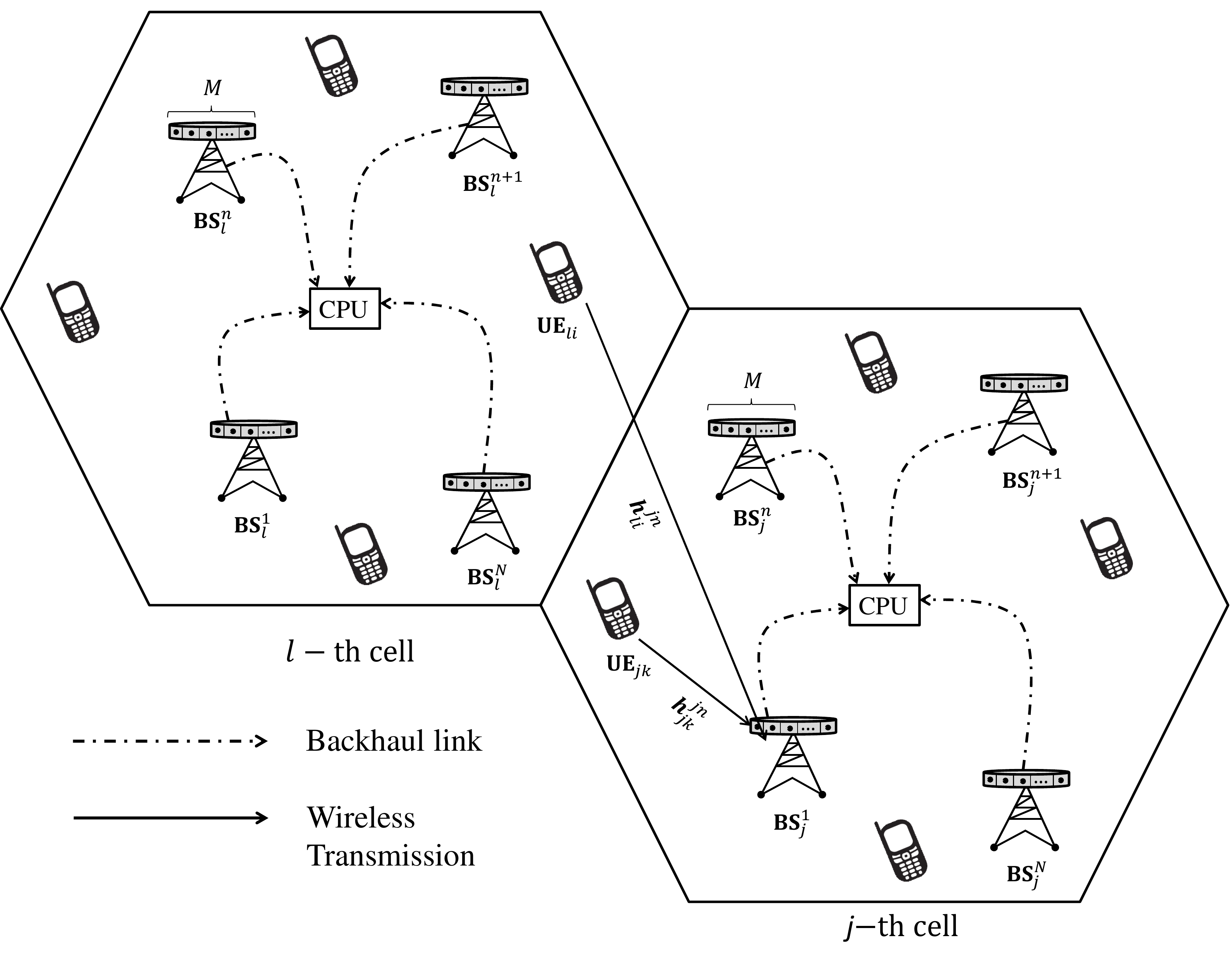}}
%\caption{DAA massive MIMO.}
\label{DAA_disp}
\end{subfigure}
\caption{An illustration of the multi-cell DAA massive MIMO system. Each DAA massive MIMO
cell operate autonomously, i.e., there is no ``network MIMO'' or ``CoMP''.}
\label{network_config}
%\end{minipage}
\end{figure}

In various practical and foreseen deployment scenarios, the antennas connected to a given BS may not be arranged in a compact array.  For example, each BS may be connected to several compact sub-arrays that serve different sectors.  Alternatively, antennas may be distributed over facades or rooftops of buildings.  Consequently, the distance-dependent path-loss may vary among the antenna elements belonging to one BS. If non-omnidirectional antennas are used at BSs, the antenna pointing direction will also influence the path-loss. Herein, we use the term ``distributed antenna array'' (DAA) to refer to a setup where the BS antennas are grouped into sub-arrays. Importantly, for a given user terminal, each sub-array sees a different path loss but the antennas within each sub-array see the same path-loss.  The case of concern is a multi-cell DAA system, comprising multiple autonomous cells, where each cell is served by a BS. Furthermore, each DAA BS is connected to multiple compact sub-arrays, as depicted in Fig.~\ref{network_config}.  Note that as a special case, a DAA system with only a single cell and each sub-array comprising of only few antennas, would be equivalent to the setup called ``cell-free massive MIMO'' in \cite{Nayebi2015,Ngo2017}.

Power control in massive MIMO is essential in order to provide uniformly good service throughout the network. Recent research works has extensively studied power control in cell-free massive MIMO systems. By jointly optimizing the number of active access points (APs) and the transmitted power by each AP, it is possible to minimize the total downlink power consumption in cell-free massive MIMO while satisfying the individual downlink rate requirements of users in the network \cite{Chien2019}. Joint power allocation and scheduling in cell-free massive MIMO was studied while considering the intermittent network data traffic \cite{Chen2019}. Similarly, \cite{Francis2019} analyzed different power control policies in cell-free massive MIMO where a subset of available APs may be utilized to serve the users. Furthermore, \cite{Zhang2019} considers the max-min power control for multicast cell-free massive MIMO. Additionally, the power allocation problem in cell-free massive MIMO can also be considered and solved as a deep-learning problem \cite{Andrea2019}. Since in a multi-cell DAA system the path-losses are different to the different same-cell DAAs, existing algorithms for power control, such as those in \cite[Chap.~5--6]{Marzetta2016}, \cite{Li2016,Cheng2017} are inapplicable. Some existing multi-cell power control algorithms take the interference caused to BSs in another cell into account, but in a conceptually different way. In the proposed DAA massive MIMO system, the signals received at the distributed BSs within the same cell are treated as desired signals instead of interfering signals. As such, the proposed system model and setup is fundamentally different than traditional massive MIMO.

In this paper, we derive and evaluate new power-control algorithms for multi-cell massive MIMO with DAA.  The specific novel contributions of our work are summarized as follows:
\begin{enumerate}
\item
We derive a generalized closed-form expression for the downlink signal-to-interference-plus-noise ratio (SINR) valid for multi-cell massive MIMO with an arbitrary number of sub-arrays per DAA massive MIMO cell, under the assumption of correlated Rayleigh channel fading. Results for uncorrelated channel fading follow as special cases.

\item
We investigate optimal downlink power allocation and equal-$\nu$ power allocation in various configurations of DAA massive MIMO. Specifically, we formulate the downlink power allocation problem as a max-min optimization problem and then rewrite it as a target SINR maximization problem, the solution to which ensures uniformly good quality of service.

\item We analyze and compare the throughputs obtained by various configurations of DAA multi-cell massive MIMO. We also demonstrate that the fading covariance significantly influences the performance.
\end{enumerate}
In addition, we present detailed numerical results to highlight and emphasize the performance advantages achieved by DAA massive MIMO. Notably, for some scenarios, DAA massive MIMO is capable of boosting the network performance by up to $55\%$. Furthermore, we reveal that under the assumption of uncorrelated channel fading, increasing DAAs and reducing the number of antennas per sub-array may not always provide an improvement in performance.

The system model considered in our work is novel and has major distinctions as compared to the cell-free massive MIMO systems. Importantly, the proposed system model is a multi-cell model that is general enough to encompass cell-free massive MIMO as a special case. Furthermore, the proposed system model describes a more realistic scenario where both inter-cell and intra-cell interferences are present and affect the performance of the system. Cell-free massive MIMO systems only have intra-cell interference and cannot describe the impact of inter-cell interference in a straightforward manner.

This paper is a comprehensive extension of our conference paper \cite{Akbar2018}. The main novel contributions include investigating correlated fading and providing new insightful numerical results. In \cite{Akbar2018}, we confirmed the performance benefits obtained by DAA massive MIMO. Building on this preliminary study, this paper provides a comprehensive performance analysis for our proposed power control algorithm in various configurations of DAA massive MIMO. In \cite{Akbar2018}, the users' locations were assumed to be the same for numerical results. Differently, in this paper we consider various user locations and evaluate the median and $95\%$ likely performances, which reliably indicates the network performance. Additionally, in this paper we examine the network performance under the assumptions of both correlated and uncorrelated channel fading, which is an important advancement relative to \cite{Akbar2018}.

\textit{Notations:} We denote vectors and matrices by lower-case boldface symbols and upper-case boldface symbols, respectively. $\mathbb{E}[\cdot]$ denotes the expectation, $\|\cdot\|$ denotes the $\textit{l}_2$ norm, $\textrm{tr}(\cdot)$ denotes the matrix trace, $(\cdot)^{\textrm{H}}$ denotes the Hermitian transpose, and $(\cdot)^{T}$ denotes the matrix transpose, and $\mathbf{I}_M$ denotes an $M\times M$ identity matrix.

\section{System Model}
In this paper, we consider a multi-cell multi-user massive MIMO network, as illustrated in Fig.~\ref{network_config}. The network consists of $L$ cells with $K$ single-antenna users in each cell\footnote{We consider the traditional definition of cells, where a coverage area is divided into non-overlapping regions and the users in each region is served by the network infrastructure located in the region. Different from traditional cellular networks where each cell is served by one BS, our network infrastructure in a cell consists of multiple distributed BSs.}. Notably, each cell has $N$ DAAs deployed at arbitrary locations, where each sub-array is equipped with $M$ antenna elements. As such, there are $M_{\textrm{tot}}=M\times N$ antenna elements in each cell. Each sub-array in a cell is connected to a CPU through back-haul links \footnote{In DAA massive MIMO, there is an information exchange overhead between sub-arrays and the CPU. The impact of information exchange overhead on the performance of the DAA massive MIMO is beyond the scope of the current work. A recent study analyzed the exchange of information between multiple CPUs in a cell-free system and demonstrated that it is possible to obtain performance similar to a single-CPU cell-free system \cite{Palou2019}. Furthermore, we assume that the information exchange between the sub-arrays and the CPU is perfectly synchronized.}. We highlight that no coordination is required between different cells in the system and CPU in each cell only requires local information via a backhaul links that can be implemented using cloud-RAN techniques. Furthermore, we assume that there is perfect synchronization in the system. We denote the BS in the \textit{j}-th cell by $\textrm{BS}_{j}$, where $j\in\{1,\cdots,L\}$. The \textit{n}-th antenna sub-array in the \textit{j}-th cell is represented by $\textrm{BS}_{j}^n$, where $n\in\{1,\cdots,N\}$. Furthermore, we represent the \textit{k}-th user in the \textit{j}-th cell by $\textrm{U}_{jk}$, where $k\in\{1,\cdots,K\}$. We represent the uplink channel between $\textrm{U}_{jk}$ and $\textrm{BS}_{l}^n$ by $\mathbf{h}_{jk}^{ln}$, where $l\in\{1,\cdots,L\}$. Additionally, we assume that the channels follow a correlated Rayleigh fading distribution, i.e., $\mathbf{h}_{jk}^{ln} \sim \mathcal{CN}(\mathbf{0},\mathbf{R}_{jk}^{ln})$, where $\mathbf{R}_{jk}^{ln}$ is the channel covariance matrix, which encapsulates various channel impairments such as average path-loss and spatial correlation. We clarify that the path loss between a user and all antenna elements of a DAA is considered constant. However, due to the physical separation between various DAAs the path losses between a users and different DAAs in a cell are not constant. As such, the existing power control algorithm cannot be applied to the system model considered in our work. The system model considered in this paper is a generalized model as compared to \cite{Nguyen2017,Nayebi2015,Ngo2017} since it is capable of describing various antenna array deployments. For example, we note that the cell free massive MIMO is a special case of our considered system model when $L=1$, $M=1$, and $N=M_{\textrm{tot}}$. Moreover, co-located massive MIMO is another special case of our considered system model when $M=M_{\textrm{tot}}$ and $N=1$.

We assume the network operates in the time division duplex (TDD) mode. Accordingly, the uplink and downlink channels are assumed to be the same and reciprocal within one channel coherence interval \cite{Marzetta2016}. Consequently, the BS utilizes the uplink channel estimates for downlink precoding based on the assumption of channel reciprocity. In the beginning of each channel coherence interval, the users in each cell transmit their pilot sequences to the same-cell BS, which then performs channel estimation. The channel estimation is followed by the downlink data transmission where each BS sends data to the same-cell users.

\subsection{Uplink Channel Estimation Under Perfect Channel Knowledge}
At the beginning of each channel coherence interval, all users send their pre-assigned pilot sequences to the same-cell BS for the purpose of channel estimation. We denote the pilot sequence assigned to $\textrm{U}_{jk}$ by $\mathbf{\boldsymbol{\phi}}_{jk}$, where $\|\mathbf{\boldsymbol{\phi}}_{jk}\|^2=1$, $j\in\{1,\cdots,L\}$, and $k\in\{1,\cdots,K\}$. We assume that each pilot sequence is of length $\tau_p$. We highlight that we need to estimate the channels between a user and all the same-cell DAAs in the system. Differently, in co-located massive MIMO, the channel estimation is typically performed between a user and a single same-cell BS. Furthermore, we assume that each user within the same cell is assigned an orthogonal pilot sequence, i.e., $\tau_p=K$. The same set of pilot sequences are reused in each cell across the entire network. Consequently, the uplink pilot transmission received at the \textit{n}-th sub-array of $\textrm{BS}_{j}$, i.e., $\textrm{BS}_{j}^n$, is given as
\begin{align} \label{uplink_trans}
\mathbf{Y}^{jn} &= \sum\limits_{l=1}^L\sum\limits_{i=1}^K\mathbf{h}_{li}^{jn}\mathbf{\boldsymbol{\phi}}_{li}^\textrm{H} + \frac{1}{\sqrt{\rho_{\textrm{tr}}}}\mathbf{N}_{j}^n,
\end{align}
where $\mathbf{N}_{j}^n\in \mathbf{\mathbb{C}}^{M \times \tau}$ represents the additive white Gaussian noise (AWGN) at $\textrm{BS}_{j}^n$, and $\rho_{\textrm{tr}}$ is the normalized pilot power per user. Afterwards, the sub-array $\textrm{BS}_{j}^n$ correlates \eqref{uplink_trans} with the known pilot sequence to obtain
\begin{align} \label{uplink_trans1}
\mathbf{y}_{jk}^{jn} &= \left(\sum\limits_{l=1}^L\sum\limits_{i=1}^K\mathbf{h}_{li}^{jn}\mathbf{\boldsymbol{\phi}}_{li}^\textrm{H} + \frac{1}{\sqrt{\rho_{\textrm{tr}}}}\mathbf{N}_{j}^n\right)\mathbf{\boldsymbol{\phi}}_{jk}.
\end{align}
We denote the correlation between pilot sequences assigned to $\textrm{U}_{li}$ and $\textrm{U}_{jk}$ as $\rho_{li}^{jk}=\mathbf{\boldsymbol{\phi}}_{li}^\textrm{H}\mathbf{\boldsymbol{\phi}}_{jk}$. Based on this definition, we re-express \eqref{uplink_trans1} as
\begin{align} \label{uplink_trans2}
\mathbf{y}_{jk}^{jn} &= \mathbf{h}_{jk}^{jn} + \sum\limits_{\substack{l=1 \\ (l,i) \neq}}^L \sum\limits_{\substack{i=1 \\ (j,k)}}^K \rho_{li}^{jk}\mathbf{h}_{li}^{jn} + \frac{1}{\sqrt{\rho_{\textrm{tr}}}}\mathbf{N}_{j}^n\mathbf{\boldsymbol{\phi}}_{jk}.
\end{align}
From \eqref{uplink_trans2}, we obtain the MMSE uplink channel estimate, i.e., $\mathbf{\widehat{h}}_{jk}^{jn}$ of the channel $\mathbf{h}_{jk}^{jn}$ as \cite{Emil2016a}
\begin{align} \label{mmse_est}
\mathbf{\widehat{h}}_{jk}^{jn} &= \mathbf{W}_{jk}^{n}\mathbf{y}_{jk}^{jn},
\end{align}
where
\begin{align}\label{W_val}
\mathbf{W}_{jk}^{n}=\mathbf{R}_{jk}^{jn}(\mathbf{Q}_{jk}^{n})^{-1},
\end{align}
\begin{align}\label{R_val}
\mathbf{R}_{jk}^{jn}=\mathbb{E}\left[\mathbf{h}_{jk}^{jn}(\mathbf{h}_{jk}^{jn})^\textrm{H}\right],
\end{align}
and
\begin{align}\label{Q_val}
  \mathbf{Q}_{jk}^{n}&=\mathbb{E}\left[\mathbf{y}_{jk}^{jn}(\mathbf{y}_{jk}^{jn})^\textrm{H}\right]= \sum\limits_{l=1}^L \sum\limits_{i=1}^K \left|\rho_{li}^{jk}\right|^2 \mathbf{R}_{lk}^{jn} + \frac{1}{\rho_{\textrm{tr}}}\mathbf{I}_M.
\end{align}
Under the assumption of full statistical channel knowledge, the covariance matrices $\mathbf{R}_{jk}^{jn}$ are known to the BSs. As such, $\textrm{BS}_{j}^n$  can obtain the matrix $\mathbf{Q}_{jk}^n$ by using \eqref{Q_val}. Afterwards, utilizing \eqref{W_val} together with \eqref{uplink_trans2}, $\textrm{BS}_{j}^n$  obtain the channel uplink channel estimates using \eqref{mmse_est}. We highlight that the assumption of perfect channel covariance knowledge is commonly used in massive MIMO literature \cite{Emil2016a,Yin2013,Adhikary2013}. Additionally, the change in the channel covariance information occurs at a slow rate \cite{Adhikary2013}. As such, the channel statistics remains largely unchanged over several channel coherence intervals. Furthermore, several methods exist in literature to estimate the change in the channel covariance information with small overhead \cite{Marzetta2011,Hoydis2011}.

\subsection{Uplink Channel Estimation Under Limited Channel Covariance Knowledge}
In this subsection, we discuss a more practical scenario where BSs only have limited knowledge of the channel covariance. Specifically, we assume that BSs do not have full knowledge of the channel covariance matrix $\mathbf{R}_{jk}^{jn}$. Furthermore, BSs only have knowledge about the diagonal elements of the channel covariance matrices. Under these assumptions, we obtain the element-wise EW-MMSE uplink channel estimate of the $z$-th element of $\mathbf{h}_{jk}^{jn}$, where $z\in\{1,\dotsc,M\}$, as \cite{Hoydis2011}
\begin{align} \label{locha}
\left[\mathbf{\widehat{\bar{h}}}{}_{jk}^{jn}\right]_z &= \frac{\left[\mathbf{R}_{jk}^{jn}\right]_{z}}{\sum_{l=1}^{L}\sum_{i=1}^{K}\left|\rho_{li}^{jk}\right|^2\left[\mathbf{R}_{li}^{jn}\right]_{z} + \frac{1}{\rho_{\textrm{tr}}}} \left[\mathbf{y}_{jk}^{jn}\right]_{z}.
\end{align}
We highlight that the diagonal elements of the channel covariance matrices are easy to estimate and require only a few additional resources \cite{Emil2016a}.
Consequently, the EW-MMSE channel estimate for $\mathbf{h}_{jk}^{jn}$ is obtained as \cite{Hoydis2011}
\begin{align} \label{mmse_est2}
\mathbf{\widehat{\bar{h}}}{}_{jk}^{jn} &= \mathbf{\widehat{W}}_{jk}^{n}\mathbf{y}_{jk}^{jn},
\end{align}
where
\begin{align}\label{W_val2}
\mathbf{\widehat{W}}_{jk}^{n}=\mathbf{D}_{jk}^{jn}(\bm{\Lambda}_{jk}^{n})^{-1}.
\end{align}
We highlight that $\mathbf{D}_{jk}^{jn}$ and $\bm{\Lambda}_{jk}^{jn}$ are $M\times M$ diagonal matrices. We define $\mathbf{D}_{jk}^{jn}$ as
\begin{align}\label{R_val2}
\mathbf{D}_{jk}^{jn} =
\begin{bmatrix}
\left[\mathbf{R}_{jk}^{jn}\right]_{1} & 0 & 0 & 0\\
0 & \left[\mathbf{R}_{jk}^{jn}\right]_{2} & 0 & 0\\
0 & 0 & ... & 0\\
0 & 0 & 0 & \left[\mathbf{R}_{jk}^{jn}\right]_{M}
\end{bmatrix}
\end{align}
and $\mathbf{\bm{\Lambda}}_{jk}^{n}$ as
\begin{align}\label{Q_val2}
\begin{bmatrix}
\sum\limits_{l,i}\left|\rho_{li}^{jk}\right|^2\left[\mathbf{R}_{li}^{jn}\right]_1 + \frac{1}{\rho_{\textrm{tr}}} & 0 & 0\\
0 & ... & 0\\
0 & 0 & \sum\limits_{l,i}\left|\rho_{li}^{jk}\right|^2\left[\mathbf{R}_{li}^{jn}\right]_M + \frac{1}{\rho_{\textrm{tr}}}
\end{bmatrix}.
\end{align}
Afterwards, utilizing \eqref{W_val2} together with \eqref{uplink_trans2}, $\textrm{BS}_{j}^n$  obtains the channel uplink channel estimates according to \eqref{mmse_est2}, under the assumption of imperfect channel knowledge at BSs.

\subsection{Downlink Data Transmission}
During the downlink data transmission phase, $\textrm{BS}_{j}$ transmits data symbols to each user in the \textit{j}-th cell. Based on the assumption of channel reciprocity, the channel estimates obtained through the uplink channel estimation are utilized in downlink transmission. Accordingly, the symbol transmitted by $\textrm{BS}_j$ for the $K$ same-cell users is represented as
\begin{align}\label{symbol_n}
  x_{j} &= \sum_{i=1}^{K}\sum_{n=1}^{N} \nu_{ji}^n \mathbf{{a}}_{ji}^{n}q_{ji},
\end{align}
where $\nu_{ji}^n \geq 0$ is the real-valued downlink power control coefficient for $\textrm{U}_{ji}$ at $\textrm{BS}_{j}^n$, $\mathbf{{a}}_{ji}^{n}$ is the downlink precoding vector for $\textrm{U}_{ji}$ at $\textrm{BS}_{j}^n$, $q_{ji}$ is the data symbol intended for $\textrm{U}_{ji}$, and ${q}_{ji} \sim \mathcal{CN}\left(0,1\right)$. We assume that the downlink power control coefficients have some constraints and are chosen to satisfy $\mathbb{E}\left[|x_{j}|^2\right] \leq 1$. Using \eqref{symbol_n}, we simplify this power constraint and represent it as
\begin{align}\label{pow_constaint}
\sum_{i=1}^K \sum_{n=1}^{N} ({\nu_{ji}^n})^2\mathbb{E}\left[\|\mathbf{{a}}_{ji}^{n}\|^2\right] &\leq 1,~\forall~j.
\end{align}

We highlight that the constraint in \eqref{pow_constaint} represents the total transmit power constraint in cell $j$, which is normalized such that the maximum power is 1. From \eqref{symbol_n}, the downlink signal received at $\textrm{U}_{jk}$ is
\begin{align}\label{pow_constaint1}
r_{jk} &=\sum_{l=1}^L\sum_{i=1}^K\sum_{n=1}^N \nu_{li}^n (\mathbf{h}_{jk}^{ln})^\textrm{H} \mathbf{{a}}_{li}^{n}q_{li}+ n_{jk},
\end{align}
where $n_{jk}$ is the AWGN at $\textrm{U}_{jk}$. Assuming that the users only have the statistical information about the channels, and the instantaneous channel information is not available at the users due to the lack of downlink pilots \cite{Akbar2016a}, we represent the downlink signal received at $\textrm{U}_{jk}$ as
\begin{align} \label{received1}
r_{jk} &=\sum\limits_{n=1}^N \nu_{jk}^n \mathbb{E}\left[(\mathbf{h}_{jk}^{jn})^\textrm{H} \mathbf{{a}}_{jk}^{n}\right]q_{jk} + \sum_{\substack{l,i,n \\(l,i) \neq (j,k)}}^{L,K,N} \nu_{li}^n(\mathbf{h}_{jk}^{ln})^\textrm{H} \mathbf{{a}}_{li}^{n}q_{li} \nonumber \\ & +\sum\limits_{n=1}^N \nu_{jk}^n\bigg((\mathbf{h}_{jk}^{jn})^\textrm{H}\mathbf{{a}}_{jk}^{n} - \mathbb{E}\left[(\mathbf{h}_{jk}^{jn})^\textrm{H}\mathbf{{a}}_{jk}^{n}\right]\bigg)q_{jk} + n_{jk}.
\end{align}
\normalsize
We highlight that the first term on the right hand side (RHS) of \eqref{received1} denotes the $N$ superimposed copies of the downlink data symbol $q_{jk}$ received from the $N$ sub-arrays in cell $j$. We highlight that \eqref{received1} is a generalized expression for received signal at $\textrm{U}_{jk}$, which is valid for an arbitrary number of DAAs in a cell.

\subsection{Achievable Downlink Rate and Throughput}
In this subsection, we derive a closed-form expression for the downlink rate. Based on the derived expression, we obtain the downlink throughput.

We note that the last three terms on RHS of \eqref{received1} are the effective noise. We also note that, these three terms are uncorrelated with the first term in \eqref{received1}. Accordingly, the downlink rate for $\textrm{U}_{jk}$ is given as \cite{Emil2017bb}
\begin{align}\label{SE_chan}
{R}_{jk}&=\log_{2}\left(1+\gamma_{jk}\right)& \textrm{b/s/Hz},
\end{align}
where $\gamma_{jk}$ is the effective downlink SINR for $\textrm{U}_{jk}$ given by \eqref{SINR_chan} at the top of the next page. We obtain the downlink throughput achieved by $\textrm{U}_{jk}$ as
\begin{figure*}[!t]
\begin{align} \label{SINR_chan}
\gamma_{jk}=\frac{\left| \textstyle{\sum_{n=1}^N} \nu_{jk}^n \mathbb{E}\left[(\mathbf{h}_{jk}^{jn})^\textrm{H} \mathbf{{a}}_{jk}^{n} \right]\right|^2}{\textstyle{\sum_{l=1}^L\sum_{i=1}^K} \mathbb{E}\left[\left|\sum_{n=1}^N\nu_{li}^n(\mathbf{h}_{jk}^{ln})^\textrm{H} \mathbf{{a}}_{li}^{n}\right|^2\right] - \left|\textstyle{\sum_{n=1}^N} \nu_{jk}^n\mathbb{E}\left[(\mathbf{h}_{jk}^{jn})^\textrm{H} \mathbf{{a}}_{jk}^{n}\right]\right|^2 + \sigma_{n}^2}. %\tag{19}
\end{align}
\setcounter{equation}{19}
\begin{align} \label{SINR}
{\gamma}_{jk}^{\textrm{MMSE}}=\frac{\left|\textstyle{\sum_{n=1}^N}\textrm{tr}\left(\nu_{jk}^n\mathbf{W}_{jk}^{n}\mathbf{R}_{jk}^{jn}\right)\right|^2} {\textstyle{\sum_{l=1}^L \sum_{i=1}^K\sum_{n=1}^N} \textrm{tr}\left((\nu_{li}^n)^2\mathbf{W}_{li}^n \mathbf{Q}_{li}^n(\mathbf{W}_{li}^n)^\textrm{H}\mathbf{R}_{jk}^{ln}\right) + \textstyle{\sum_{l=1,l\neq j}^L} \left|\sum_{n=1}^N\textrm{tr}\left(\nu_{lk}^n\mathbf{W}_{lk}^n\mathbf{R}_{jk}^{ln}\right)\right|^2 + \sigma_{n}^2}. %\tag{15}
\end{align}
\setcounter{equation}{21}
\begin{align} \label{SINR_EW}
{\gamma}_{jk}^{\textrm{EW-MMSE}}=\frac{\left|\textstyle{\sum_{n=1}^N}\textrm{tr}\left(\nu_{jk}^n\mathbf{\widehat{W}}_{jk}^{n} \mathbf{R}_{jk}^{jn}\right)\right|^2} {\textstyle{\sum_{l=1}^L \sum_{i=1}^K\sum_{n=1}^N} \textrm{tr}\left((\nu_{li}^n)^2\mathbf{\widehat{W}}_{li}^n \mathbf{Q}_{li}^n(\mathbf{\widehat{W}}_{li}^n)^\textrm{H}\mathbf{R}_{jk}^{ln}\right) + \textstyle{\sum_{l=1,l\neq j}^L} \left|\sum_{n=1}^N\textrm{tr}\left(\nu_{lk}^n\mathbf{\widehat{W}}_{lk}^n\mathbf{R}_{jk}^{ln}\right)\right|^2 + \sigma_{n}^2}. %\tag{17}
\end{align}
\hrulefill
\vspace*{4pt}
\end{figure*}
\setcounter{equation}{18}
\begin{align}\label{tp_chan2}
\textrm{Throughput}_{jk}& = \textrm{BW} \times \ell \times {R}_{jk} & \textrm{b/s},
\end{align}
where $\textrm{BW}$ denotes that channel bandwidth and $\ell$ denotes the portion of coherence interval used for the downlink data transmission. In this paper, we use the average per-user throughput as the performance metric for comparison, which is based on the numerical average obtained from individual user throughputs. We highlight that if all the users in the network achieve the same downlink SINR, the average per-user throughput is the same as the individual user throughput. We next provide a closed-form expression for the SINR in the following theorem.
\begin{theorem} \label{theorem}
Assuming that the BSs perform maximum ratio transmission (MRT) in the downlink, i.e, $\mathbf{a}_{jk}^{n} = \mathbf{\widehat{h}}_{jk}^{jn}$, the closed-form expression for the downlink effective SINR at $\textrm{U}_{jk}$ is obtained as in \eqref{SINR} at the top of the next page.
\end{theorem}
\begin{IEEEproof}
The proof is given in Appendix \ref{SINR_proof}.
\end{IEEEproof}
The closed-form expression for the downlink SINR given in \eqref{SINR} can be re-written as
\setcounter{equation}{20}
\begin{align}\label{SINR_reduced}
{\gamma}_{jk}^{\textrm{MMSE}}=\frac{\left|\textstyle{\sum_{n=1}^N}\nu_{jk}^n\chi_{jk}^{n}\right|^2} {\textstyle{\sum_{l,i,n}^{L,K,N}}(\nu_{li}^n)^2\zeta_{jk}^{lin} + \textstyle{\sum_{l\neq j}^{L}}|\textstyle{\sum_{n=1}^N}\nu_{lk}^n\xi_{jk}^{ln}|^2 + \sigma_{n}^2},
\end{align}
where
  $\chi_{jk}^{n} = \textrm{tr}(\mathbf{W}_{jk}^{n}\mathbf{R}_{jk}^{jn})$,
  $\zeta_{jk}^{lin} = \textrm{tr}(\mathbf{W}_{li}^n \mathbf{Q}_{li}^n(\mathbf{W}_{li}^n)^\textrm{H}\mathbf{R}_{jk}^{ln})$, and
  $\xi_{jk}^{ln} = \textrm{tr}(\mathbf{W}_{lk}^n\mathbf{R}_{jk}^{ln})$.
\begin{theorem} \label{theorem2}
Assuming that the BSs perform maximum ratio transmission (MRT) in the downlink, i.e, $\mathbf{a}_{jk}^{n} = \mathbf{\widehat{\bar{h}}}_{jk}^{jn}$, the closed-form expression for the downlink effective SINR at $\textrm{U}_{jk}$ is obtained as in \eqref{SINR_EW} at the top of the page.
\end{theorem}
\begin{IEEEproof}
The proof is given in Appendix \ref{SINR_proof2}.
\end{IEEEproof}
The closed-form expression for the downlink SINR given in \eqref{SINR_EW} can be re-written as
\setcounter{equation}{22}
\begin{align}\label{SINR_reduced2}
{\gamma}_{jk}^{\textrm{EW-MMSE}}=\frac{\left|\textstyle{\sum_{n=1}^N}\nu_{jk}^n\bar{\chi}_{jk}^{n}\right|^2} {\textstyle{\sum_{l,i,n}^{L,K,N}}(\nu_{li}^n)^2\bar{\zeta}_{jk}^{lin} + \textstyle{\sum_{l\neq j}^{L}}|\textstyle{\sum_{n=1}^N}\nu_{lk}^n\bar{\xi}_{jk}^{ln}|^2 + \sigma_{n}^2},
\end{align}
where
  $\bar{\chi}_{jk}^{n} = \textrm{tr}(\mathbf{\widehat{W}}_{jk}^{n}\mathbf{R}_{jk}^{jn})$,
  $\bar{\zeta}_{jk}^{lin} = \textrm{tr}(\mathbf{\widehat{W}}_{li}^n \mathbf{Q}_{li}^n(\mathbf{\widehat{W}}_{li}^n)^\textrm{H}\mathbf{R}_{jk}^{ln})$, and
  $\bar{\xi}_{jk}^{ln} = \textrm{tr}(\mathbf{W}_{lk}^n\mathbf{R}_{jk}^{ln})$.

\section{Downlink Power Control in Distributed Antenna Array Massive MIMO}
In this section, we formulate the downlink power control problem as a max-min optimization problem. Max-min power control maximizes the minimum rate or SINR for all the user in the network. As such, every user in the network receives a uniform quality of service. Max-min power control has previously been studied for conventional co-located BSs \cite{Yang2014,Marzetta2016}. However, the application of max-min power control for DAA massive MIMO networks, where each array has multiple antenna elements, has not been investigated.

The proposed algorithm only requires the estimated channel information between users and the same-cell DAAs and no coordination is required between different cells. Importantly, in TDD operation BSs can listen to the transmissions from other cells and estimate all the parameters that are needed. As such, it is possible to estimate the inter-cell parameters using the same procedure as the intra-cell parameters. Accordingly, additional signaling is not required. The goal of the max-min optimization problem is to maximize the minimum downlink SINR for all the users in the network. As such, we formulate the network-wide max-min optimization problem using \eqref{SINR_reduced} as
\begin{align} \label{opt_problem}
\begin{aligned}
& \underset{\{\nu_{li}^n\}}{\text{max}}~~~\underset{\forall~j,k}{\text{min}}
& & \frac{\left|\textstyle{\sum_{n=1}^N}\nu_{jk}^n\chi_{jk}^{n}\right|^2} {\textstyle{\sum_{l,i,n}^{L,K,N}}(\nu_{li}^n)^2\zeta_{jk}^{lin} + \textstyle{\sum_{l\neq j}^{L}}|\textstyle{\sum_{n=1}^N}\nu_{lk}^n\xi_{jk}^{ln}|^2 + \sigma_{n}^2} \\
& \text{s. t.}
& & \hspace{-0.4cm} \textstyle{\sum_{i=1}^K\textstyle{\sum_{n=1}^N}(\nu_{li}^n)^2\textrm{tr}(\mathbf{W}_{li}^n \mathbf{Q}_{li}^n\left(\mathbf{W}_{li}^n\right)^\textrm{H}) \leq 1},\forall~l,n,\\
& & & \hspace{-0.4cm} \nu_{li}^n \geq 0, \;\forall~l,i,n,
\end{aligned}
\end{align}
where the constraint $\sum_{i=1}^K\sum_{n=1}^N(\nu_{li}^n)^2\textrm{tr}(\mathbf{W}_{li}^n \mathbf{Q}_{li}^n\left(\mathbf{W}_{li}^n\right)^\textrm{H}) \leq 1$ is obtained from \eqref{pow_constaint} under the assumption that MRT is used at the BSs. The first constraint in the optimization problem (19) establishes that the normalized sum of all the power control coefficients does not exceed $1$. The second constraint ensures that the power control coefficients are non-negative. Assuming that the target SINR is ${\gamma}$, we rewrite the optimization problem given in \eqref{opt_problem} on the epigraph form as
\begin{align} \label{opt_problem1}
\begin{aligned}
& \underset{\{\nu_{li}^n\},{\gamma}}{\text{max}}
& & \gamma \\
& \text{s. t.}
& & \hspace{-0.25cm} \frac{\left|\textstyle{\sum_{n=1}^N}\nu_{jk}^n\chi_{jk}^{n}\right|^2} {\textstyle{\sum_{l,i,n}^{L,K,N}}(\nu_{li}^n)^2\zeta_{jk}^{lin} + \textstyle{\sum_{l\neq j}^{L}}|\textstyle{\sum_{n=1}^N}\nu_{lk}^n\xi_{jk}^{ln}|^2 + \sigma_{n}^2} \geq {\gamma} \\
& & &\hspace{-0.25cm} \forall\;j,k, \\
& & & \hspace{-0.25cm} \textstyle{\sum_{i=1}^K\textstyle{\sum_{n=1}^N}(\nu_{li}^n)^2\textrm{tr}(\mathbf{W}_{li}^n \mathbf{Q}_{li}^n\left(\mathbf{W}_{li}^n\right)^\textrm{H}) \leq 1},\forall~l,n,\\
& & &\hspace{-0.25cm} \nu_{li}^n \geq 0,\; \forall~l,i,n.
\end{aligned}
\end{align}
This problem can be solved as a quasi-convex program. We next formulate a convex feasibility problem based on \eqref{opt_problem1}, which we use in a bisection algorithm \cite{Boyd2004} to search for the value of $\gamma\in[\gamma_{\textrm{min}},\gamma_{\textrm{max}}]$ that is the global optimum to \eqref{opt_problem1}, where ${\gamma}_{\textrm{min}}$ and ${\gamma}_{\textrm{max}}$ define the search range \cite{Ngo2017,Boyd2004}.
\begin{proposition} \label{prop_1}
The constraint set in the optimization problem \eqref{opt_problem1} is convex and the optimization problem is quasi-concave. For a given constant value of $\gamma$, the optimization problem in \eqref{opt_problem1} is re-written as the convex feasibility problem\footnote{We highlight that \eqref{opt_problem2} is a general convex feasibility problem where the objective function is zero. Furthermore, \eqref{opt_problem1} can be considered as a special case of the general convex feasibility problem \eqref{opt_problem2}.}
\begin{align} \label{opt_problem2}
\begin{aligned}
& \underset{\{\nu_{li}^n\}}{\text{max}}
& & 0 \\
& \text{s. t.}
& &   \|\mathbf{x}_{jk}\| \leq \frac{1}{\sqrt{\gamma}}{\left|\textstyle{\sum_{n=1}^N}\nu_{jk}^n\chi_{jk}^{n}\right|},\; \forall~j,k, \\
%& &  &\textrm{tr}(\mathbf{W}_{jk}^{n}\mathbf{R}_{jk}^{jn}) \leq \chi_{jk}^{n}, \forall~n,\\
%& & &\textrm{tr}(\mathbf{W}_{li}^n \mathbf{Q}_{li}^n\left(\mathbf{W}_{li}^n\right)^\textrm{H}\mathbf{R}_{jk}^{ln}) \leq \zeta_{jk}^{lin} ,\;\forall~l,i,n\\
%& & &\textrm{tr}(\mathbf{W}_{lk}^n\mathbf{R}_{jk}^{ln})\leq \xi_{jk}^{ln},\;\forall~l,n\\
%& & & \nu_{lk}^n\xi_{jk}^{ln} \leq \varrho_{jk}^{lin}, \forall~l,n,\\
& & & \textstyle{\sum_{i=1}^K\sum_{n=1}^N(\nu_{li}^n)^2\textrm{tr}(\mathbf{W}_{li}^n \mathbf{Q}_{li}^n\left(\mathbf{W}_{li}^n\right)^\textrm{H}) \leq 1},\; \forall~l,\\
& & &\nu_{li}^n \geq 0,\;\forall\;l,i,n,
\end{aligned}
\end{align}
where $\mathbf{x}_{jk} = [\mathbf{\tilde{x}}_{jk}~~\mathbf{\bar{x}}_{jk}~~\sqrt{\sigma_{n}^2}]^T$. We define $\mathbf{\tilde{x}}_{jk}$ and $\mathbf{\bar{x}}_{jk}$ as $\mathbf{\tilde{x}}_{jk}=[\mathbf{\tilde{x}}_{jk}^{11}\dotsc\mathbf{\tilde{x}}_{jk}^{li}\dotsc\mathbf{\tilde{x}}_{jk}^{LK}]$ and $\mathbf{\bar{x}}_{jk}=[\mathbf{\bar{x}}_{jk}^{11}\dotsc\mathbf{\bar{x}}_{jk}^{li}\dotsc\mathbf{\bar{x}}_{jk}^{LK}]$, respectively, where
\begin{align}\label{def_x11}
\mathbf{\tilde{x}}_{jk}^{li}=[(\nu_{li}^1)(\zeta_{jk}^{li1})^{\frac{1}{2}}\dotsc(\nu_{li}^N)(\zeta_{jk}^{liN})^{\frac{1}{2}}]
\end{align}
and
\begin{align}\label{def_x11}
\mathbf{\bar{x}}_{jk}^{li}=\begin{cases}
\varrho_{jk}^{lk1}+\varrho_{jk}^{lkn}+\dotsc+\varrho_{jk}^{lkN} & l\neq j,\\
{0}, & l=j.
\end{cases}
\end{align}
\end{proposition}
\begin{IEEEproof}
The proof is given in Appendix \ref{SOCP_proof}.
\end{IEEEproof}

We next outline the step-by-step approach to solve the convex feasibility problem (21) as follows:
\begin{enumerate}
  \item In each iteration of the bisection algorithm, we set $\bar{\gamma}=({\gamma}_{\textrm{min}}+{\gamma}_{\textrm{max}})/2$ and solve the feasibility problem \eqref{opt_problem2} by setting ${\gamma}=\bar{\gamma}$.
  \item If the problem is infeasible, we set ${\gamma}_{\textrm{max}}=\bar{\gamma}$ otherwise we set ${\gamma}_{\textrm{min}}=\bar{\gamma}$.
  \item The algorithm iteratively refines ${\gamma}_{\textrm{min}}$ and ${\gamma}_{\textrm{max}}$ and stops the search when ${\gamma}_{\textrm{max}}-{\gamma}_{\textrm{max}}<\varepsilon$, where $\varepsilon>0$ is the error tolerance.
\end{enumerate}
To summarize, the bisection algorithm modifies $\gamma_{\textrm{min}}$ and $\gamma_{\textrm{min}}$ based on certain conditions to find a value for $\gamma$, where the optimization problem \eqref{opt_problem2} is optimal. We highlight that the max-min power control in cell-free massive MIMO \cite{Nayebi2015} is a special case of the power control problem considered in this paper, which can be obtained from \eqref{opt_problem2} when the network has one cell and each antenna array has one or more antenna elements.

The number of iterations required to solve the bisection algorithm is given as $\log_2( \gamma_{max} - \gamma_{min}) - \log_2(\varepsilon)$ \cite{emilbook}. As such, the algorithm has low computational complexity and converges very fast, since the search range is halved in each iteration. We highlight that a second-order cone program (SOCP) is solved in each iteration. The computational complexity of the SOCP is $\mathcal{O}(K_{\textrm{tot}}^4)$, where $K_{\textrm{tot}}=LK$ \cite{Lobo1998,bashar2018}. The total number of arithmetic operations required to solve the optimization problem \eqref{opt_problem2} is given as $[\log_2( \gamma_{max} - \gamma_{min}) - \log_2(\varepsilon)] \times \mathcal{O}(K_{\textrm{tot}}^4)$.

The max-min power control in \eqref{opt_problem1} maximizes the minimum SINR. In this work, we evaluate the average per-user throughput to demonstrate the performance of the power control algorithms. We highlight that the average per-user throughput is dependent on the effective SINR as given in \eqref{SE_chan} and \eqref{tp_chan2}. Therefore, by evaluating the average per-user throughput, we also demonstrate the improvement in SINR. Furthermore, we highlight that using the procedure outlined in this subsection, it is possible to obtain the max-min power control for \eqref{SINR_reduced2}.

\subsection{Equal-$\nu$ Power Allocation}
In this paper, we use equal-$\nu$ power allocation as a baseline for comparison with the proposed power control. In equal-$\nu$ power allocation, the total available downlink transmit power is shared equally among all the users in a cell \cite{Yang2014}. As such, the downlink power control coefficients ${\nu_{li}^n}$ are equal for all the users. From \eqref{pow_constaint} and assuming that the full available power is used by the BSs during the downlink transmission, we obtain
\begin{align}\label{pow_constaint_equal1}
(\nu)^2\sum_{l=1}^L\sum_{i=1}^K \sum_{n=1}^{N} \mathbb{E}\left[\|\mathbf{{a}}_{li}^{n}\|^2\right] &= L.
\end{align}
Assuming that the BSs performs MRT, the power control coefficient is obtained as
\begin{align}\label{equal_pow}
  {\nu} &= \sqrt{\frac{L}{\sum_{l=1}^L\sum_{i=1}^K \sum_{n=1}^{N}\textrm{tr}\left(\mathbf{W}_{li}^n \mathbf{Q}_{li}^n\left(\mathbf{W}_{li}^n\right)^\textrm{H}\right)}}.
\end{align}
Intuitively, equal-$\nu$ power allocation implies that every user is allocated power from a serving antenna array proportionally to the mean-square of its channel estimate. As such, a user receives more power from the arrays with good propagation conditions than the arrays with weaker propagation conditions.

With equal-$\nu$ power allocation, the power control coefficients $\nu$ remain the same regardless of channel conditions. Accordingly, it is expected that equal-$\nu$ power allocation does not give a higher throughput as compared to the max-min power control. However, equal-$\nu$ allocation serves as an important benchmark for the network performance.

\section{Numerical Results}
In this section, we numerically demonstrate the performance benefits of DAA massive MIMO. Throughout the section, we assume that $L=2$ and there are $K=10$ users in each cell. We consider hexagonal cells with radius $1000$ m. Each user within a cell is allocated an orthogonal pilot sequence and the same pilot sequence set is repeated in the neighbouring cell \footnote{Let us assume that the channel coherence time is $T_c=3\;\textrm{ms}$ and the coherence bandwidth is $W_c=300\;\textrm{kHz}$, the coherence block length in this case is $S=T_c W_c = 900$ symbols. If we assume that $40\%$ of the coherence block length is used for the uplink pilot training, we have $360$ orthogonal pilot sequences available in a cell. As such, orthogonal pilot sequences can be allocate to $360$ users within a cell, which is a reasonably large number of users. As such, it is reasonable to assume that all the users within a cell can be allocated orthogonal pilot sequences.}. Additionally, we assume that there are $M_{\textrm{tot}}=400$ antenna elements in each cell. The CDFs are obtained from the average per-user throughput for $100$ different realization of user locations, where the users' locations are uniformly distributed within a cell. Furthermore, the sub-arrays are deployed in multiple tiers at an offset distance of $120$ m from the center of cell. We assume that $20\;\textrm{MHz}$ bandwidth is available for transmission and $45\%$ of the coherence interval is used for the downlink transmission. Furthermore, all the cells in the network have the same geometry. We use CVX \cite{CVX} to solve the max-min power control problem. For the sake of clarity, we represent a DAA massive MIMO having $N$ antenna arrays and $M$ antennas per sub-array as $\{M,N\}$ DAA massive MIMO. We obtain the median per-user throughput by computing the median of the average per-user throughputs obtained for $100$ different realization of the user locations. The path loss model is $1/d^\kappa$, where $d$ is the distance between a user and a sub-array and $\kappa=3.76$. In Section V.A and Section V.B, we assume that the BSs have perfect channel knowledge. In Section V.C, we assume that the BSs have limited channel knowledge. The simulation parameters remain the same unless stated otherwise.

\subsection{Max-Min and Equal-$\nu$ Power Allocation under Uncorrelated Channel Fading}
In this subsection, we examine the impact of max-min power allocation and equal-$\nu$ power allocation in DAA massive MIMO under the assumption of uncorrelated channel fading.
\begin{figure}[!t]
\centering
\includegraphics[width=3.4in]{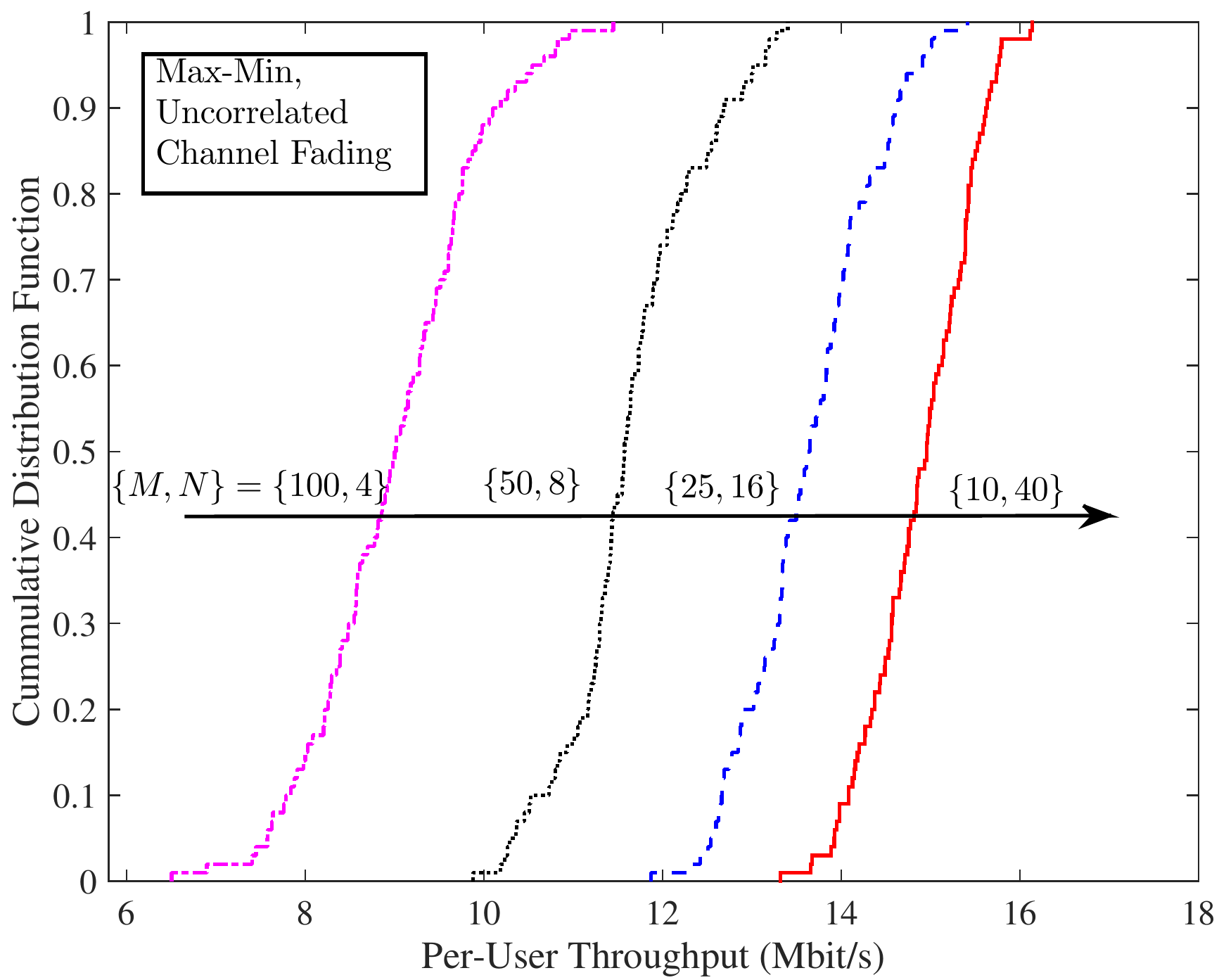}
\caption{The CDFs of max-min power allocation under uncorrelated channel fading for $\{100,4\}$, $\{50,8\}$, $\{25,16\}$, and $\{10,40\}$ DAA configuration.} \label{MaxminUncorrelated100to10}
\end{figure}

We first analyze the throughput achieved by applying max-min power control in DAA massive MIMO. Fig.~\ref{MaxminUncorrelated100to10} depicts the CDFs of max-min power allocation for $\{100,4\}$, $\{50,8\}$, $\{25,16\}$, and $\{10,40\}$ DAA massive MIMO. We note that DAA massive MIMO provides a large improvement in the per-user throughput. Importantly, as $N$ increases, there is an increase in the per-user throughput. For example, when $N$ increases from $4$ to $40$, the median per-user throughput increases from $11.6$ Mbit/s to $15.0$ Mbit/s, which is equivalent to $22.7\%$ increase. Similarly, when $N$ increases from $4$ to $40$, the $95\%$ likely performance increases from $7.6$ Mbit/s to $14.0$ Mbit/s, which corresponds to an $45.6\%$ improvement in the per-user throughput. Furthermore, we note that the individual throughput values are closer to the median value when $N=40$. Differently, the spread of the individual throughput values around the median is larger for $N=4$. As such, in the $\{10,40\}$ DAA configuration, the majority of the users enjoy a higher throughput close to the median throughput value.

\begin{figure}[!t]
\centering
\includegraphics[width=3.4in]{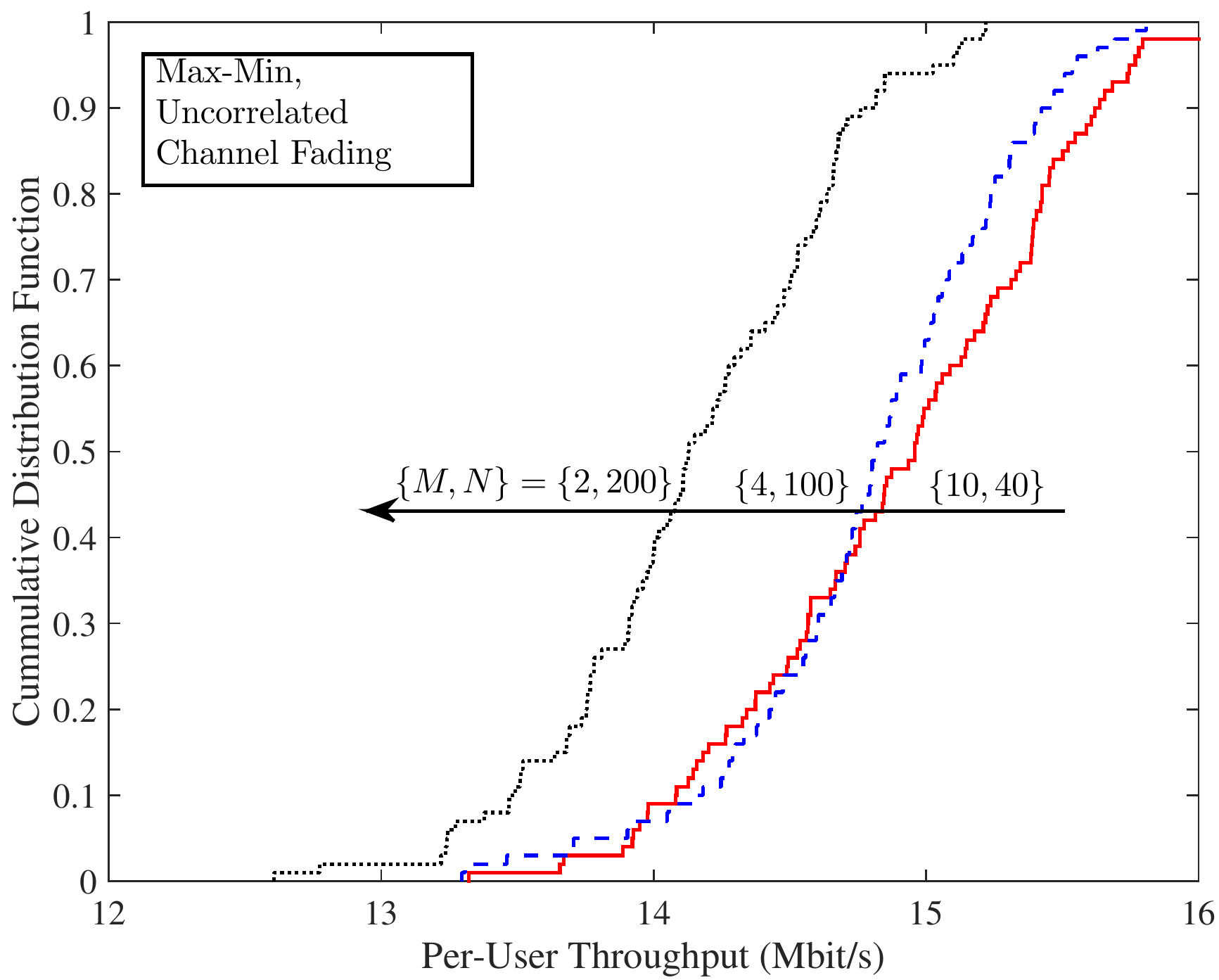}
\caption{The CDFs of max-min power allocation under uncorrelated channel fading for $\{10,40\}$, $\{4,100\}$, and $\{2,200\}$ DAA configuration.} \label{MaxminUncorrelated10to2}
\end{figure}

We next examine the impact of decreasing the number of antennas in each sub-array to a small value, i.e., below $M=10$. Fig.~\ref{MaxminUncorrelated10to2} depicts the CDFs of the per-user throughput for $\{10,40\}$, $\{4,100\}$, and $\{2,200\}$ DAA massive MIMO. In contrast to the results obtained in Fig.~\ref{MaxminUncorrelated100to10}, we observe that increasing $N$ above $40$ and decreasing $M$ below $10$ reduces the per-user throughput. For example, when $M$ is reduced from $10$ to $2$, the median per-user throughput decreases from $15.0$ Mbit/s to $14.1$ Mbit/s, which corresponds to a $5.5\%$ decrease. Similarly, when $M$ decreases from $10$ to $2$, the $95\%$ likely performance drops from $13.9$ Mbit/s to $13.2$ Mbit/s or equivalently $4.9\%$. As such, we observe a $5\%$ loss in performance by reducing the number of antenna elements per sub-array below $10$. We highlight that this behaviour is due to a loss in channel hardening \cite{Chen2018}. Specifically, reducing $M$ deteriorate the benefits offered by favourable propagation in massive MIMO as a result of increased uncertainty in channel statistics. Furthermore, we highlight that in our simulations the exact $M$ where we observe a degradation in performance is between $M=4$ and $M=10$.

\begin{figure}[!t]
\centering
\includegraphics[width=3.4in]{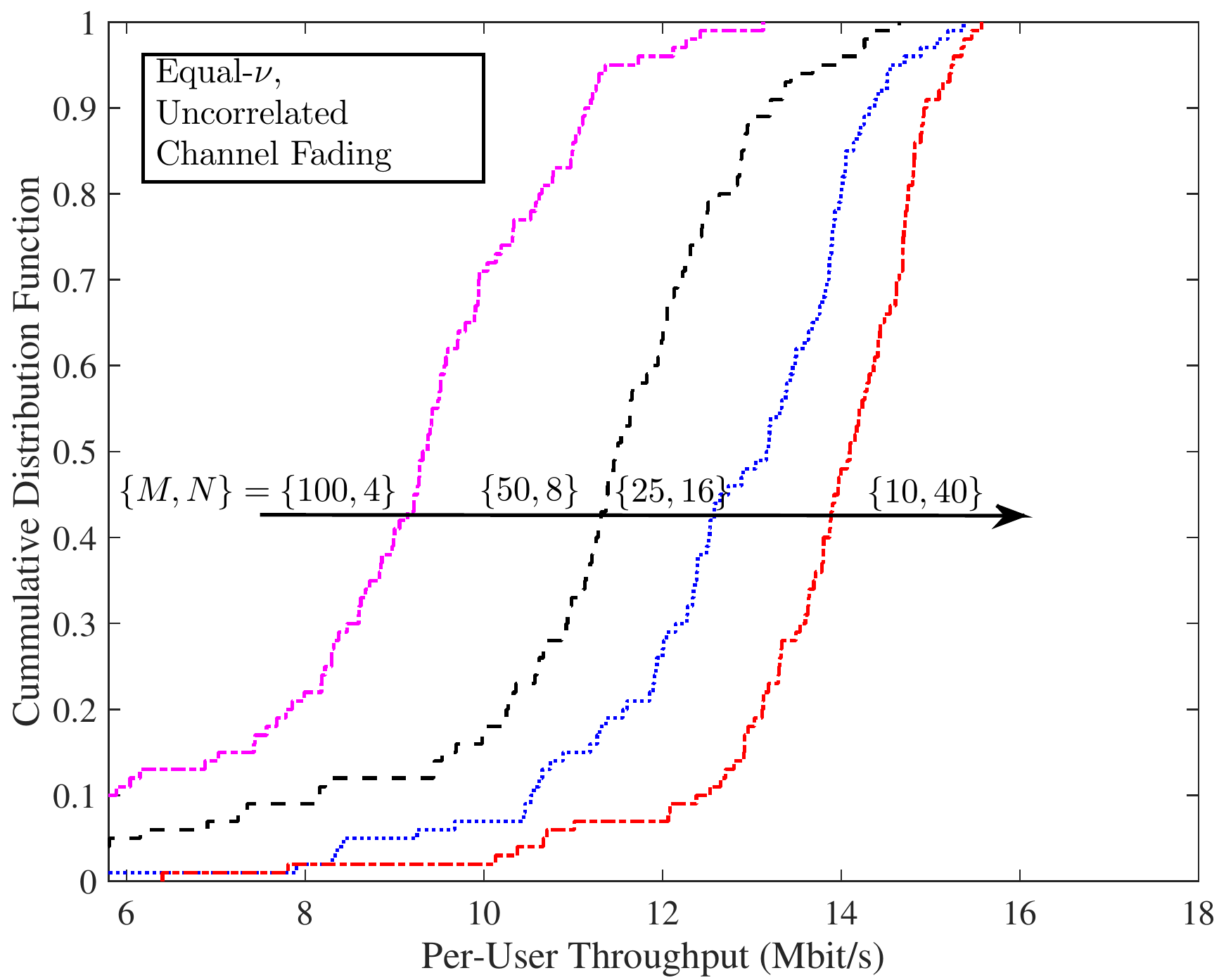}
\caption{The CDFs of equal-$\nu$ power allocation under uncorrelated channel fading for $\{100,4\}$, $\{50,8\}$, $\{25,16\}$, and $\{10,40\}$ DAA configuration.} \label{EqualUncorrelated100to10}
\end{figure}

We now analyze equal-$\nu$ power allocation in DAA massive MIMO. Fig.~\ref{EqualUncorrelated100to10} shows the CDFs of the per-user throughput for various configurations of DAA massive MIMO. We highlight that equal-$\nu$ power control achieves a lower per-user throughput as compared to max-min power control for the same DAA configuration. For example, for $\{10,40\}$ DAA massive MIMO, equal-$\nu$ power control achieves $14.10$ Mbit/s median per-user throughput and $10.7$ Mbit/s $95\%$ likely performance, which is $5.8\%$ and $23.0\%$, respectively, lower than that achieved by max-min power control for the same DAA configuration. As such, when $N$ increases from $4$ to $40$, the median per-user throughput increases from $11.5$ Mbit/s to $14.1$ Mbit/s (or equivalently, by $18.5\%$), and the $95\%$ likely performance increases from $4.8$ Mbit/s to $10.7$ Mbit/s (or equivalently, by $55.4\%$).

\begin{figure}[!t]
\centering
\includegraphics[width=3.4in]{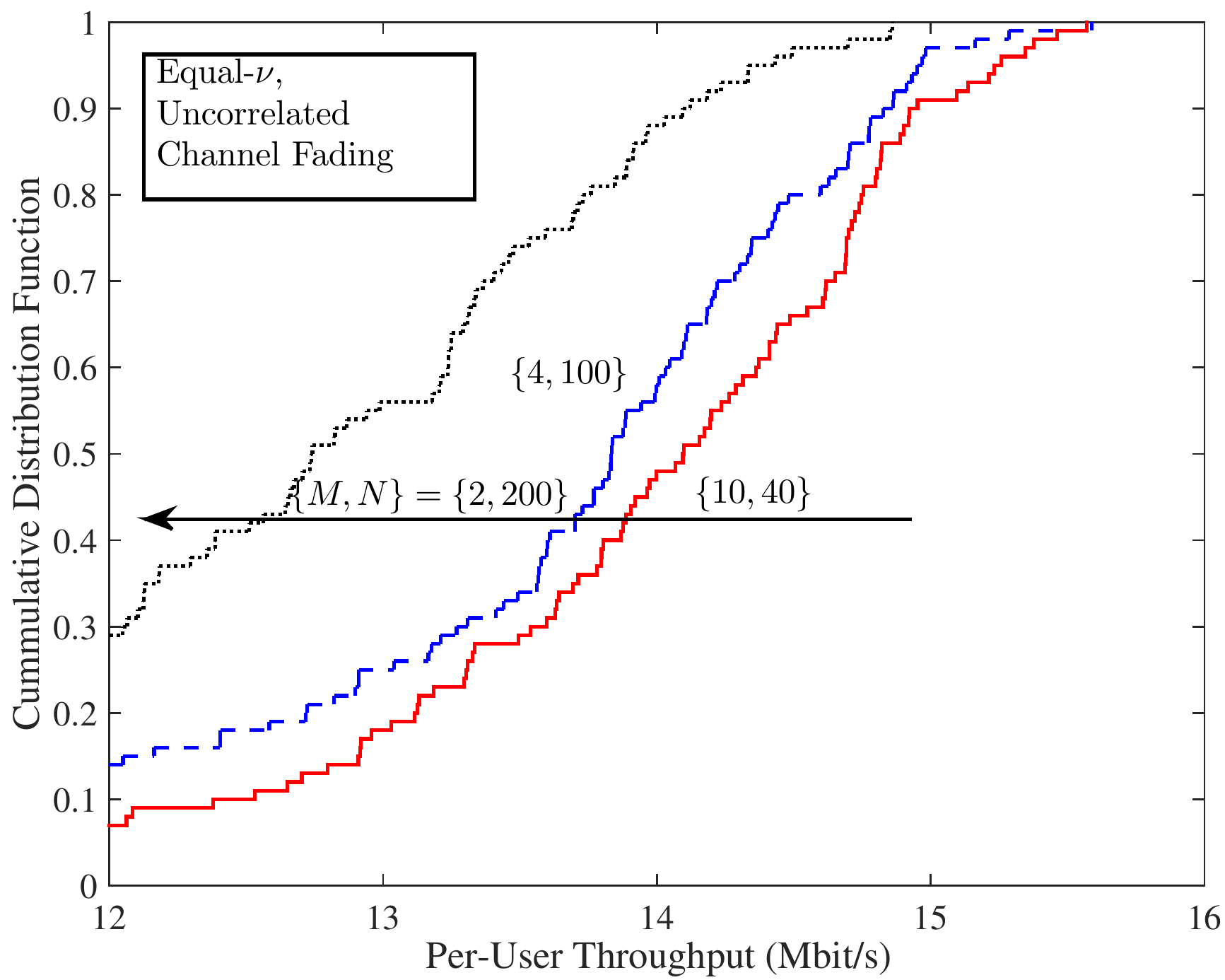}
\caption{The CDFs of equal-$\nu$ power allocation under uncorrelated channel fading for $\{10,40\}$, $\{4,100\}$, and $\{2,200\}$ DAA configuration.}\label{EqualUncorrelated10to2}
\end{figure}

Fig.~\ref{EqualUncorrelated10to2} depicts the CDFs of the per-user throughput for equal-$\nu$ power allocation when each sub-array is equipped with $M=10$ or less antenna elements. Similar to our observations in Fig.~\ref{MaxminUncorrelated10to2}, we note that reducing $M$ below $10$ degrades the median per-user throughput and the $95\%$ likely performance. For example, when $M$ reduces from $10$ to $2$, the median per-user throughput decreases from $14.1$ Mbit/s to $12.7$ Mbit/s (or equivalently, by $9.7\%$). Additionally, for the same reduction in $M$, the $95\%$ likely performance reduces from $10.7$ Mbit/s to $9.2$ Mbit/s (or equivalently, by $13.8\%$).

We highlight that under the assumption of uncorrelated channel fading, DAA massive MIMO improves the per-user throughput up to a certain value of $\{M,N\}$. In our simulations, when $M$ is below 10, we observe a loss in the achievable per-user throughput as compared to the maximum throughput achieved when $M=10$. We highlight that this behaviour is due to the significant loss in channel hardening when the number of antenna elements per sub-array is below 10 \cite{Chen2018}.

\begin{table*}[!t]
\centering
\caption{A Summary of the Throughputs (Mbit/s) for various DAA Configurations}
\label{table_simulation}
\centering
\scalebox{1.2}{\begin{tabular}{|c|c|c|c||c|c|} \hline
\textbf{Power Allocation} &\textbf{DAA Configuration} & \multicolumn{2}{c||}{\textbf{Uncorrelated Fading}} & \multicolumn{2}{c|}{\textbf{Correlated Fading}} \\ \hline \hline
                          &                           & Median & $95\%$ Likely Performance                 & Median & $95\%$ Likely Performance \\\hline
                          \multicolumn{6}{c}{}  \\\hline \hline
\multirow{3}{2.5cm}{Max-Min}  & $(100,4)$             & $11.6$  & $7.6$   &  $10.3$ & $9.3$      \\\hhline{|~-----|}
                          & $(50,8)$                  & $13.6$  & $10.4$  &  $10.5$ & $9.6$    \\\hhline{|~-----|}
                          & $(25,16)$                 & $14.2$  & $12.6$  &  $11.9$ & $10.9$  \\\hhline{|~-----|}
                          & $(10,40)$                 & \cellcolor{blue!15} $\mathbf{15.0}$  & \cellcolor{blue!15} $\mathbf{13.9}$   & $11.9$  & $11.0$\\\hhline{|~-----|}
                          & $(4,100)$                 & $14.8$  & $13.9$  &  $12.0$ & $11.2$  \\\hhline{|~-----|}
                          & $(2,200)$                 & $14.1$  & $13.2$  &  \cellcolor{blue!15} $\mathbf{12.9}$ &  \cellcolor{blue!15} $\textbf{11.9}$   \\\hline \hline
                                    \multicolumn{6}{c}{}  \\\hline \hline
\multirow{3}{2.5cm}{Equal-$\nu$}& $(100,4)$           & $11.5$  & $4.8$                                  & $7.8$  & $5.0$      \\\hhline{|~-----|}
                          & $(50,8)$                  & $13.2$  & $6.2$                                  & $9.0$  & $6.6$    \\\hhline{|~-----|}
                          & $(25,16)$                 & ${13.5}$                & $9.3$                  & $10.0$  & $6.6$  \\\hhline{|~-----|}
                          & $(10,40)$                 & \cellcolor{blue!15} $\mathbf{14.1}$  & \cellcolor{blue!15} $\mathbf{10.7}$  & $10.6$   & $7.1$ \\\hhline{|~-----|}
                          & $(4,100)$                 & $13.8$  & $10.7$  &  $10.6$ & \cellcolor{blue!15} $\mathbf{8.1}^{*}$  \\\hhline{|~-----|}
                          & $(2,200)$                 & $12.7$  & $9.2$  &  \cellcolor{blue!15} $\mathbf{11.3}$ & $7.2$  \\\hline \hline
                          \multicolumn{6}{l}{\scriptsize Note: The unit for all values is Mbit/s. $^{*}$ represents the outlier value.}  \\
\end{tabular}}
\end{table*}
\subsection{Max-Min and Equal-$\nu$ Power Allocation under Correlated Channel Fading}
In this subsection, we assume that the wireless channels undergo correlated channel fading. We generate the covariance matrix for the correlated channels using the one-ring channel model \cite{Adhikary2013}. We assume a uniformly distributed angular spread. The standard deviation of the angular spread is $5$ degrees and the spacing between adjacent antenna elements is $\frac{\lambda}{2}$, where $\lambda$ is the wavelength of the frequency used for transmission. Furthermore, we assume that the shadow fading is independent.

\begin{figure}[!t]
\centering
\includegraphics[width=3.4in]{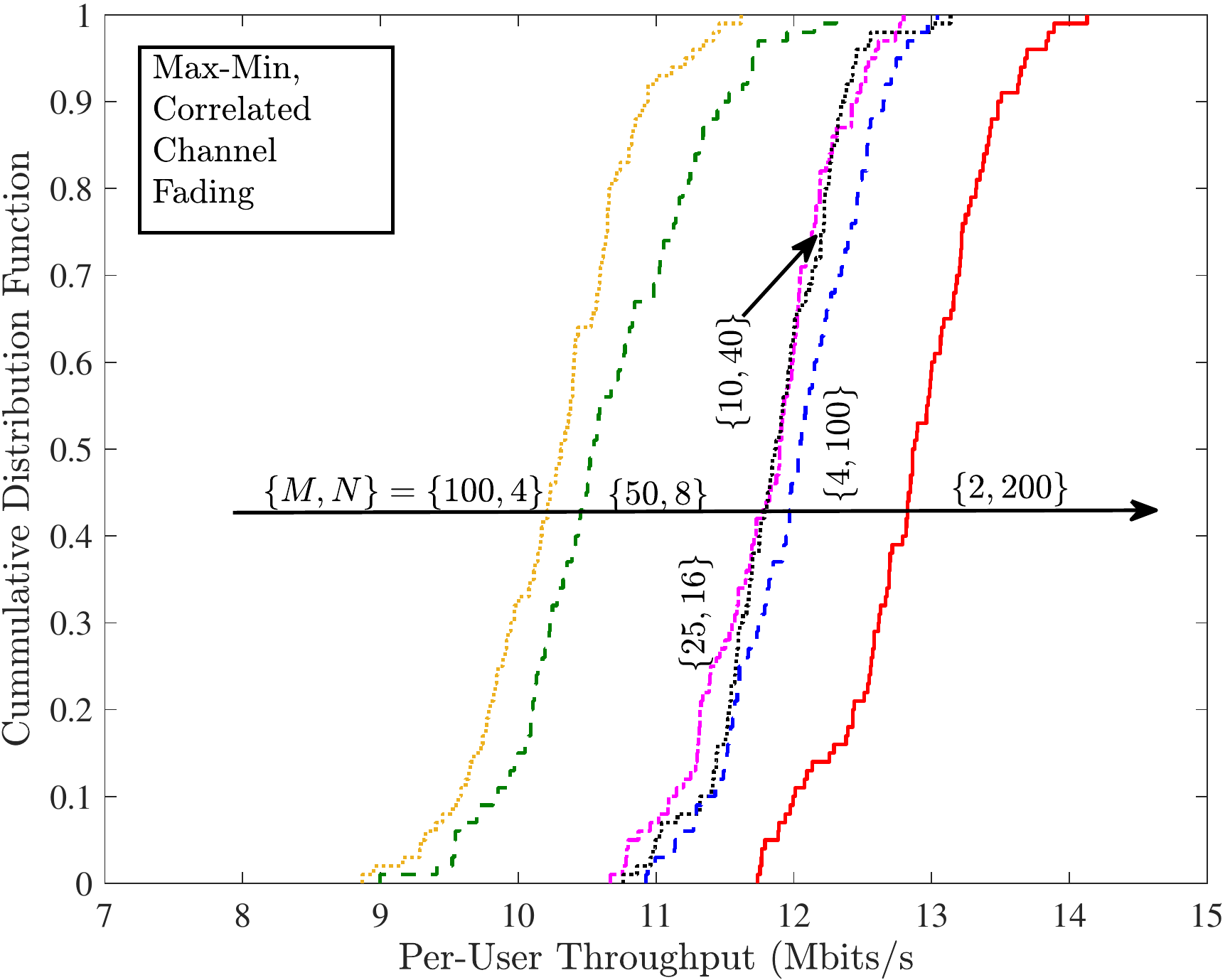}
\caption{The CDFs of max-min power allocation under correlated channel fading for $\{100,4\}$, $\{50,8\}$, $\{25,16\}$, $\{10,40\}$, $\{4,100\}$, and $\{2,200\}$ DAA configuration.} \label{MaxMinCorrelatedAll}
\end{figure}

Fig.~\ref{MaxMinCorrelatedAll} shows the CDFs of max-min power control for $\{100,4\}$, $\{50,8\}$, $\{25,16\}$, $\{10,40\}$, $\{4,100\}$, and $\{2,200\}$ DAA massive MIMO. We highlight that the median and the $95\%$ likely performance of the per-user throughput under correlated channel fading are lower than those obtained for uncorrelated channel fading for the same $N$ and $M$. For example, for $\{10,40\}$ DAA configuration, the median per-user throughput obtained from Fig.~\ref{MaxMinCorrelatedAll} is $11.9$ Mbit/s, which is $20.7\%$ lower than that obtained from Fig.~\ref{MaxminUncorrelated100to10} for the same DAA configuration. Similarly, the $95\%$ likely performance decreases to $11.0$ Mbit/s, which corresponds to a $20.8\%$ decrease as compared to uncorrelated channel fading. We observe that when $N$ increases from $4$ to $200$, the median per-user throughput increases from $10.3$ Mbit/s to $12.9$ Mbit/s (or equivalently, by $19.9\%$). Interestingly, when the channel fading is correlated, the network can deploy as little as two antenna elements per sub-array and still achieve an improvement in the throughput as compared to the network with higher $N$. Likewise, we also achieve an improvement in the $95\%$ likely performance. For instance, when $N$ increases from $4$ to $200$, the $95\%$ likely performance increases from $9.3$ Mbit/s to $11.9$ Mbit/s (or equivalently, by $21.5\%$). We highlight that, in addition to the increased throughput, max-min power control also improves the fairness in the network because each user achieves the same downlink throughput.

\begin{figure}[!t]
\centering
\includegraphics[width=3.4in]{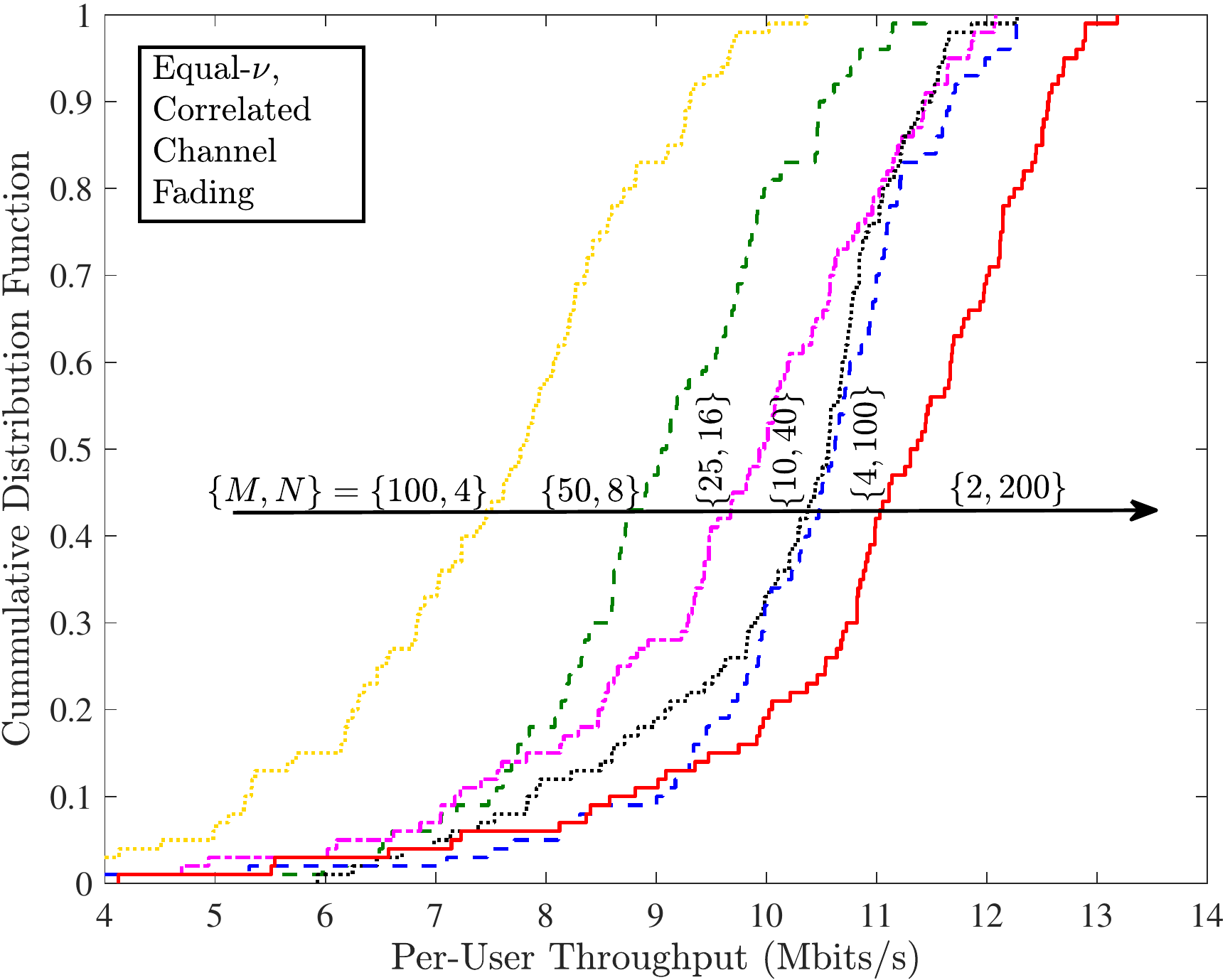}
\caption{The CDFs of equal-$\nu$ power allocation under correlated channel fading for $\{100,4\}$, $\{50,8\}$, $\{25,16\}$, $\{10,40\}$, $\{4,100\}$, and $\{2,200\}$ DAA configuration.} \label{EqualCorrelatedAll}
\end{figure}

Fig.~\ref{EqualCorrelatedAll} depicts the CDFs of equal-$\nu$ power control for various DAA configuration. We highlight that equal-$\nu$ power allocation achieves a lower per-user throughput as compared to max-min power control. Moreover, the performance of equal-$\nu$ power control in correlated channel fading is less than that in uncorrelated channel fading. For example, for $\{10,40\}$ DAA configuration, we observe $25.1\%$ reduction in the median per-user throughput and $33.4\%$ reduction in the $95\%$ likely performance as compared to uncorrelated channel fading. Furthermore, we observe that increasing $N$ from $4$ to $200$ increases the median per-user throughput by $31.4\%$. We also observe $31.1\%$ increase in the $95\%$ likely performance for the same $N$.

Table~\ref{table_simulation} provides a summary of the median and the $95\%$ likely performance for equal-$\nu$ and max-min power allocation under uncorrelated and correlated channel fading. The maximum value for each power allocation (max-min, equal-$\nu$) and fading type (correlated, uncorrelated) is highlighted in light blue colour. For the $95\%$ likely performance, we have considered the value for $\{4,100\}$ DAA configuration as an outlier. Based on the results in Table~\ref{table_simulation}, we emphasise that channel covariance is important in determining the performance of the DAA massive MIMO network. We highlight that $\{100,4\}$, which is the closest configuration to a co-located massive MIMO, provides the lowest median and $95\%$ likely performance in all the cases as evident by Table I. As such, we recommend that DAA massive MIMO should be preferred over a co-located massive MIMO regardless of the channel fading conditions (correlated or uncorrelated).

\subsection{Max-Min and Equal-$\nu$ Power Allocation under Limited Channel Covariance Knowledge}
\begin{figure}[!t]
\centering
\includegraphics[width=3.4in, height=2.9in]{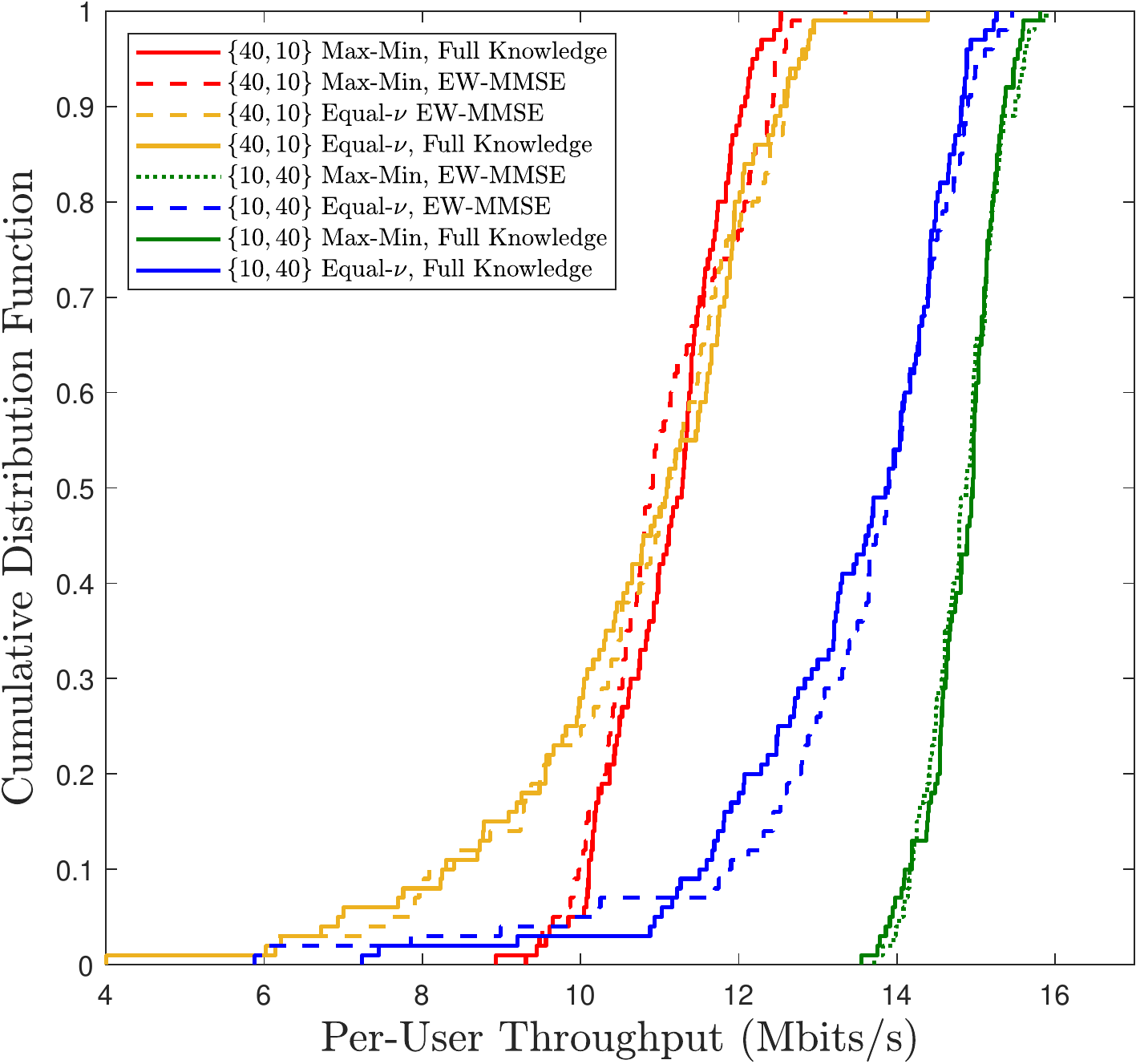}
\caption{The CDFs of max-min and equal-$\nu$ power allocation under correlated channel fading for $\{10,40\}$ and $\{40,10\}$ DAA configuration for EW-MMSE and full channel knowledge.} \label{final-EWMMSE}
\end{figure}

In this subsection, we analyze the impact of having limited channel covariance knowledge on the performance of max-min and equal-$\nu$ power allocations for $K=5$. Fig.~\ref{final-EWMMSE} depicts the CDFs of equal-$\nu$ and max-min power control under the assumption of imperfect channel knowledge and where EW-MMSE is used for channel estimation. We observe that even under the assumption of imperfect channel covariance knowledge, max-min power control is able to achieve better performance as compared to equal-$\nu$ power control. Furthermore, EW-MMSE under the assumption of imperfect channel knowledge performs reasonably well. For example, when max-min power control is used with EW-MMSE for $\{40,10\}$ configuration, the median per-user throughput is $98\%$ of the case when full channel knowledge is available. Similarly, when equal-$\nu$ power control with EW-MMSE for $\{40,10\}$ configuration is able to achieve nearly the same median per-user throughput when full channel knowledge is available to the BSs. As such, under the current system configuration and simulation parameters, the impact of having limited knowledge about channel covariance is not significant.

\section{Conclusion}
In this paper, we investigated the optimal max-min downlink power allocation in various configurations of the DAA massive MIMO network. To this end, we first derived a generalized closed-form expression for the downlink SINR under the assumption of correlated Rayleigh channel fading. Afterwards, we formulated a network-wide max-min optimization problem based on the derived closed-form downlink SINR expression. We then solved the max-min optimization problem and obtained the downlink power control coefficients. We further compared the median and $95\%$ likely performance of the optimal power allocation with equal-$\nu$ power allocation. Our numerical results indicated that adding DAAs in the network provides a large improvement in the average per-user throughput. Also, we demonstrated that the channel covariance is an important factor in determining the performance of DAA massive MIMO. Overall, the proposed system model gives network designers great flexibility to deploy BS antenna arrays in arbitrary locations and reveal how to reap the benefits offered by massive MIMO.

\appendices
\section{Proof of Theorem \ref{theorem}}\label{SINR_proof}
In this appendix, we derive the closed-form expression for the downlink SINR in \eqref{SINR_chan} for the MMSE channel estimator. The proof follows the same approach as in \cite{Emil2016a}.

We note that for MRT, the downlink precoding vector is given as $\mathbf{\widehat{a}}_{jk}^{n} = \mathbf{\widehat{h}}_{jk}^{jn}$. Accordingly, the numerator of \eqref{SINR_chan} is simplified as
\begin{align}\label{appendix1}
\mathbb{E}\left[(\mathbf{h}_{jk}^{jn})^H \mathbf{\widehat{h}}_{jk}^{jn}\right] &=\textrm{tr}(\mathbf{W}_{jk}^{n}\mathbb{E}[\mathbf{y}_{jk}^n(\mathbf{h}_{jk}^{jn})^H ]),\nonumber \\ &=
\textrm{tr}(\mathbf{W}_{jk}^{n}\mathbf{R}_{jk}^{jn}).
\end{align}
We now simplify the first term in the denominator of \eqref{SINR_chan}. We note that when $(j,k)\neq (l,i)$, $\mathbf{{h}}_{jk}^{ln}$ and $\mathbf{\widehat{h}}_{li}^{ln}$ are independent. For this case we have
\begin{align}\label{first_term}
\mathbb{E}\left[|(\mathbf{h}_{jk}^{ln})^H \mathbf{\widehat{h}}_{li}^{ln} |^2\right]  & = \textrm{tr}(\mathbf{W}_{li}^{n}\mathbf{Q}_{li}^{n}(\mathbf{W}_{li}^{n})^H\mathbf{R}_{jk}^{ln}).
\end{align}
We now consider the case where $(j,k)=(l,i)$. In this case, $\mathbf{{h}}_{jk}^{ln}$ and $\mathbf{\widehat{h}}_{li}^{ln}$ are not independent. By utilizing the fact that $\mathbf{h}_{jk}^{ln}$ and $\mathbf{\widehat{h}}_{li}^{ln}-\mathbf{W}_{li}^{n}\mathbf{h}_{jk}^{ln}$ are independent, we have
\begin{align}\label{sec_term}
\mathbb{E}\left[|(\mathbf{h}_{jk}^{ln})^H \mathbf{\widehat{h}}_{li}^{ln} |^2\right]&=  \textrm{tr}(\mathbf{W}_{li}^{n}\mathbf{Q}_{li}^{n}(\mathbf{W}_{li}^{n})^H\mathbf{R}_{jk}^{ln}) \notag \\
&+ |\textrm{tr}(\mathbf{W}_{li}^n\mathbf{R}_{jk}^{ln})|^2.
\end{align}
Combining \eqref{first_term} and the \eqref{sec_term}, the first term in the denominator of \eqref{SINR_chan} is written as
\begin{align}\label{den_first}
\mathbb{E}\left[|(\mathbf{h}_{jk}^{ln})^H \mathbf{\widehat{h}}_{li}^{ln} |^2\right]&=  \textrm{tr}(\mathbf{W}_{li}^{n}\mathbf{Q}_{li}^{n}(\mathbf{W}_{li}^{n})^H\mathbf{R}_{jk}^{ln}) + \nonumber \\
&\begin{cases}
0, & (j,k)\neq(l,i)\\
|\textrm{tr}(\mathbf{W}_{lk}^n\mathbf{R}_{jk}^{ln})|^2, & (j,k)=(l,i)
\end{cases}
\end{align}
Substituting \eqref{appendix1} and \eqref{den_first} in \eqref{SINR_chan} we obtain \eqref{SINR}, which completes the proof.

\section{Proof of Theorem \ref{theorem2}}\label{SINR_proof2}
In this appendix, we derive the closed-form expression for the downlink SINR in \eqref{SINR_chan} for the EW-MMSE channel estimator. Using \eqref{mmse_est2}, we write $\mathbf{\widehat{a}}_{jk}^{n} = \widehat{\mathbf{\bar{h}}}{}_{jk}^{jn}$. Afterwards, the numerator of \eqref{SINR_chan} is rewritten as
\begin{align}\label{appendix2}
\mathbb{E}\left[(\mathbf{h}_{jk}^{jn})^H \mathbf{\widehat{\bar{h}}}{}_{jk}^{jn}\right] &=\textrm{tr}(\mathbf{\widehat{W}}_{jk}^{n}\mathbb{E}[\mathbf{y}_{jk}^n(\mathbf{h}_{jk}^{jn})^H ]),\nonumber \\ &=
\textrm{tr}(\mathbf{\widehat{W}}_{jk}^{n}\mathbf{R}_{jk}^{jn}).
\end{align}
We next simplify the first term in the denominator of \eqref{SINR_chan}. We note that when $(j,k)\neq (l,i)$, $\mathbf{{h}}_{jk}^{ln}$ and $\mathbf{\widehat{\bar{h}}}{}_{li}^{ln}$ are independent. For this case we have
\begin{align}\label{first_term2}
\mathbb{E}\left[|(\mathbf{h}_{jk}^{ln})^H \mathbf{\widehat{\bar{h}}}{}_{li}^{ln} |^2\right]  & = \textrm{tr}(\mathbf{\widehat{W}}_{li}^{n}\mathbf{Q}_{li}^{n}(\mathbf{\widehat{W}}_{li}^{n})^H\mathbf{R}_{jk}^{ln}).
\end{align}
We now consider the case where $(j,k)=(l,i)$. In this case, $\mathbf{{h}}_{jk}^{ln}$ and $\mathbf{\widehat{\bar{h}}}{}_{li}^{ln}$ are not independent. However, by noting that $\mathbf{h}_{jk}^{ln}$ and $\mathbf{\widehat{\bar{h}}}{}_{li}^{ln}-\mathbf{\widehat{W}}_{li}^{n}\mathbf{h}_{jk}^{ln}$ are independent, we simplify the term as
\begin{align}\label{sec_term2}
\mathbb{E}\left[|(\mathbf{h}_{jk}^{ln})^H \mathbf{\widehat{\bar{h}}}{}_{li}^{ln} |^2\right]&=  \textrm{tr}(\mathbf{\widehat{W}}_{li}^{n}\mathbf{Q}_{li}^{n}(\mathbf{\widehat{W}}_{li}^{n})^H\mathbf{R}_{jk}^{ln}) \notag \\
&+ |\textrm{tr}(\mathbf{\widehat{W}}_{li}^n\mathbf{R}_{jk}^{ln})|^2.
\end{align}
Combining \eqref{first_term2} and the \eqref{sec_term2}, the first term in the denominator of \eqref{SINR_chan} is written as
\begin{align}\label{den_first2}
\mathbb{E}\left[|(\mathbf{h}_{jk}^{ln})^H \mathbf{\widehat{\bar{h}}}{}_{li}^{ln} |^2\right]&=  \textrm{tr}(\mathbf{\widehat{W}}_{li}^{n}\mathbf{Q}_{li}^{n}(\mathbf{\widehat{W}}_{li}^{n})^H\mathbf{R}_{jk}^{ln}) + \nonumber \\
&\begin{cases}
0, & (j,k)\neq(l,i)\\
|\textrm{tr}(\mathbf{\widehat{W}}_{lk}^n\mathbf{R}_{jk}^{ln})|^2, & (j,k)=(l,i)
\end{cases}
\end{align}
Substituting \eqref{appendix2} and \eqref{den_first2} in \eqref{SINR_chan} we obtain \eqref{SINR_EW}, which completes the proof.

\section{Proof of Proposition \ref{prop_1}} \label{SOCP_proof}
In this appendix, we prove that the constraint set in \eqref{opt_problem2} is convex. We write the first constraint in the optimization problem \eqref{opt_problem2} as
\begin{align} \label{socp1}
\frac{\left|\textstyle{\sum_{n=1}^N}\nu_{jk}^n\chi_{jk}^{n}\right|^2} {\textstyle{\sum_{l,i,n}^{L,K,N}}(\nu_{li}^n)^2\zeta_{jk}^{lin} + \textstyle{\sum_{l\neq j,n}^{L,N}}|\nu_{lk}^n\xi_{jk}^{ln}|^2 + \sigma_{n}^2} \geq \widehat{\gamma}_{jk} .
\end{align}
Now we introduce auxiliary variable $|\nu_{lk}^n\xi_{jk}^{ln}|^2 \leq (\varrho_{jk}^{lin})^2 $ and simplify \eqref{socp1} to obtain
\begin{align}\label{socp2}
   \left(\textstyle{\sum_{l,i,n}^{L,K,N}}(\nu_{li}^n)^2\zeta_{jk}^{lin} + \textstyle{\sum_{l\neq j,n}^{L,N}} (\varrho_{jk}^{lin})^2  + \sigma_{n}^2\right)^{\frac{1}{2}} \leq \notag \\ \frac{1}{\sqrt{\widehat{\gamma}_{jk}}}{\left|\textstyle{\sum_{n=1}^N}\nu_{jk}^n\chi_{jk}^{n}\right|},
\end{align}
which is equivalent to
\begin{align}\label{socp3}
  \|\mathbf{x}_{jk}\| &\leq \frac{1}{\sqrt{\widehat{\gamma}_{jk}}}{\left|\textstyle{\sum_{n=1}^N}\nu_{jk}^n\chi_{jk}^{n}\right|},
\end{align}
where $\mathbf{x}_{jk} = [\mathbf{\tilde{x}}_{jk}~~\mathbf{\bar{x}}_{jk}~~\sqrt{\sigma_{n}^2}]^T$. The terms $\mathbf{\tilde{x}}_{jk}$ and $\mathbf{\bar{x}}_{jk}$ are defined in \textit{Proposition 1}.

We highlight that the constraint given in \eqref{socp3} can be represented in the standard second-order-cone (SOC) form. As such, the optimization problem in \eqref{opt_problem1} is convex. Furthermore, the remaining constraints in \eqref{opt_problem1} are convex. Accordingly, the optimization problem in \eqref{opt_problem1} is quasi-concave. Accordingly, we re-write the optimization problem in \eqref{opt_problem1} as \eqref{opt_problem2} given in \textit{Proposition 1}.

\vspace{-64mm}
\begin{IEEEbiography}[{\includegraphics[width=1in,height=1.25in,clip,keepaspectratio]{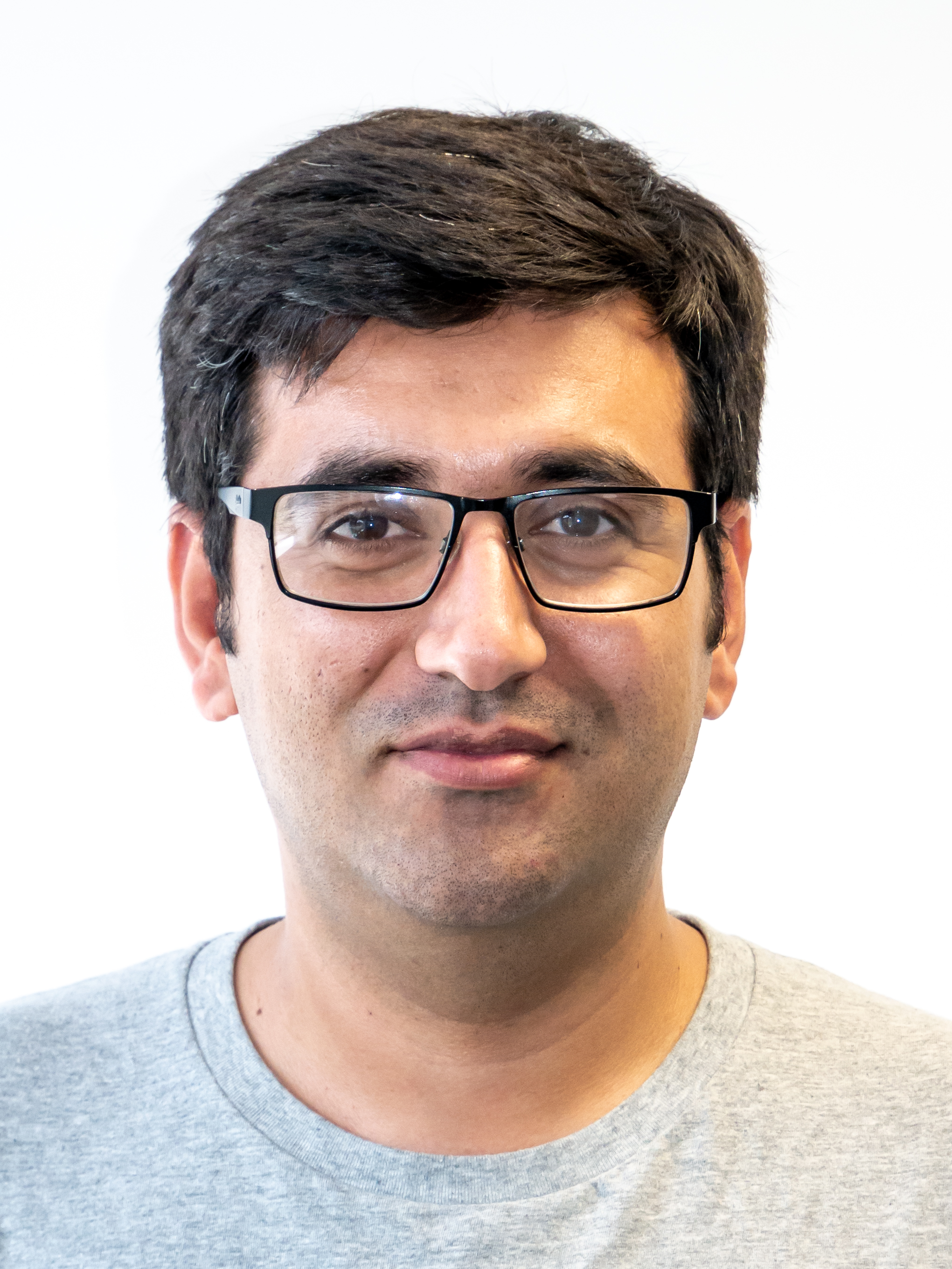}}]%
{Noman Akbar} (S’15--M'20) received the B.E. degree in electrical (telecommunication) engineering from the National University of Sciences and Technology (NUST), Rawalpindi, Pakistan, in 2012, the M.S. degree in computer engineering from Kyung Hee University, Suwon, South Korea, in 2015, and  the Ph.D. degree in electrical engineering from the Australian National University, Canberra, Australia, in 2018.  From 2012 to 2013, he was a Researcher with the NUST School of Electrical Engineering and Computer Science. From 2013 to 2015, he was with the Haptics Laboratory, Kyung Hee University, Suwon, South Korea, as a Research Assistant. He is currently a Postdoctoral Research Fellow with the Research School of Electrical, Energy and Materials Engineering, Australian National University, Canberra, Australia. He received the Best Paper Award from IEEE GlobeCOM in 2016. His research interests include signal processing for audio and wireless communications.
\end{IEEEbiography}
\vspace{-60mm}
\begin{IEEEbiography}[{\includegraphics[width=1in,height=1.25in,clip,keepaspectratio]{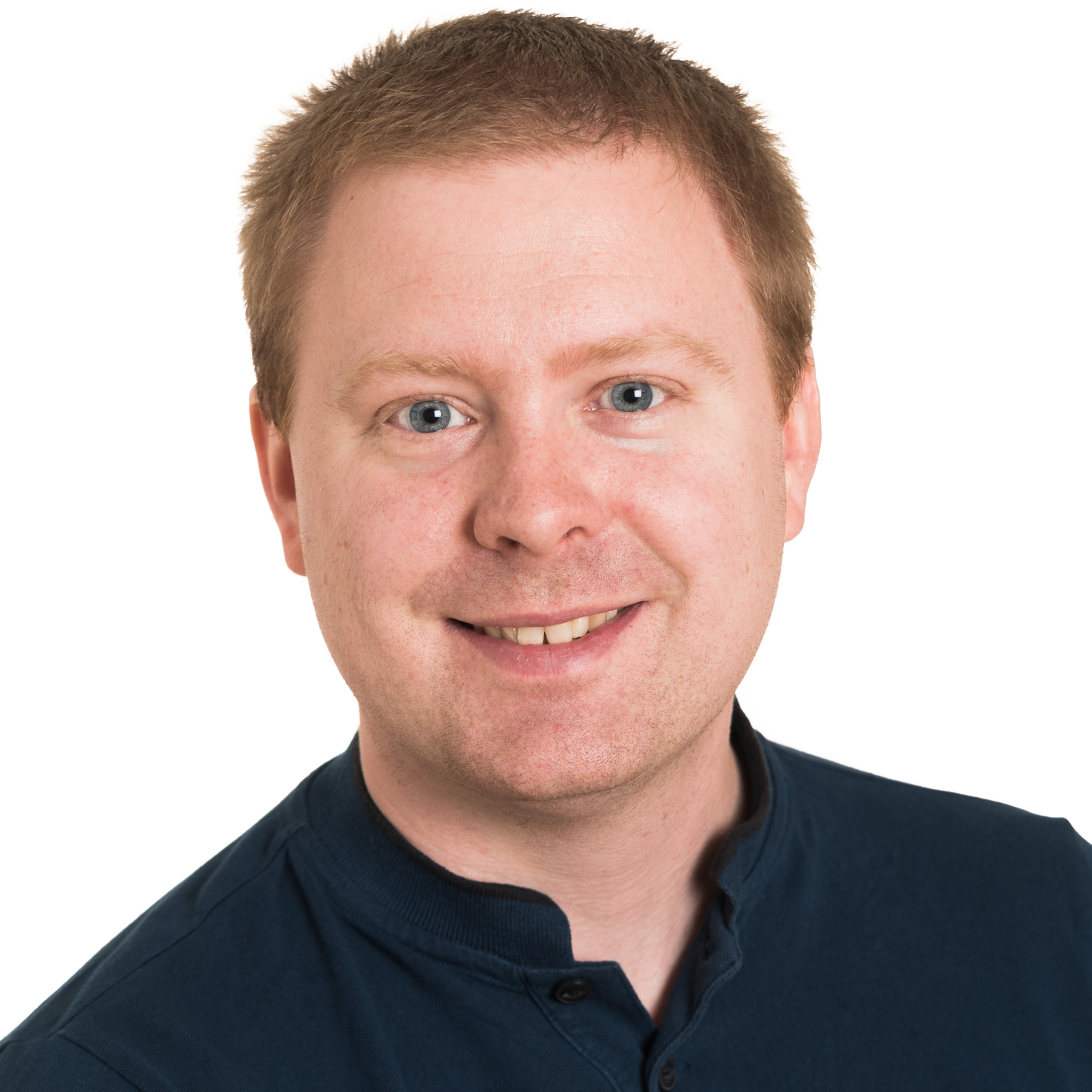}}]%
{Emil Bj\"ornson} (S'07--M'12--SM'17) received the M.S. degree in engineering mathematics from Lund University, Sweden, in 2007, and the Ph.D. degree in telecommunications from the KTH Royal Institute of Technology, Sweden, in 2011. From 2012 to 2014, he held a joint post-doctoral position at the Alcatel-Lucent Chair on Flexible Radio, SUPELEC, France, and the KTH Royal Institute of Technology. He joined Link\"oping University, Sweden, in 2014, where he is currently an Associate Professor. In September 2020, he became a part-time Visiting Full Professor at the KTH Royal Institute of Technology.

He has authored the textbooks \emph{Optimal Resource Allocation in Coordinated Multi-Cell Systems} (2013) and \emph{Massive MIMO Networks: Spectral, Energy, and Hardware Efficiency} (2017). He is dedicated to reproducible research and has made a large amount of simulation code publicly available. He performs research on MIMO communications, radio resource allocation, machine learning for communications, and energy efficiency. He has been on the Editorial Board of the IEEE Transactions on Communications since 2017. He has been a member of the Online Editorial Team of the IEEE Transactions on Wireless Communications since 2020. He has also been a guest editor of multiple special issues.

He has performed MIMO research for over ten years, his papers have received more than 10000 citations, and he has filed more than twenty patent applications. He has received the 2014 Outstanding Young Researcher Award from IEEE ComSoc EMEA, the 2015 Ingvar Carlsson Award, the 2016 Best Ph.D. Award from EURASIP, the 2018 IEEE Marconi Prize Paper Award in Wireless Communications, the 2019 EURASIP Early Career Award, the 2019 IEEE Communications Society Fred W. Ellersick Prize, and the 2019 IEEE Signal Processing Magazine Best Column Award. He also co-authored papers that received Best Paper Awards at the conferences, including WCSP 2009, the IEEE CAMSAP 2011, the IEEE SAM 2014, the IEEE WCNC 2014, the IEEE ICC 2015, and WCSP 2017.
\end{IEEEbiography}
\newpage
\begin{IEEEbiography}[{\includegraphics[width=1in,height=1.25in,clip,keepaspectratio]{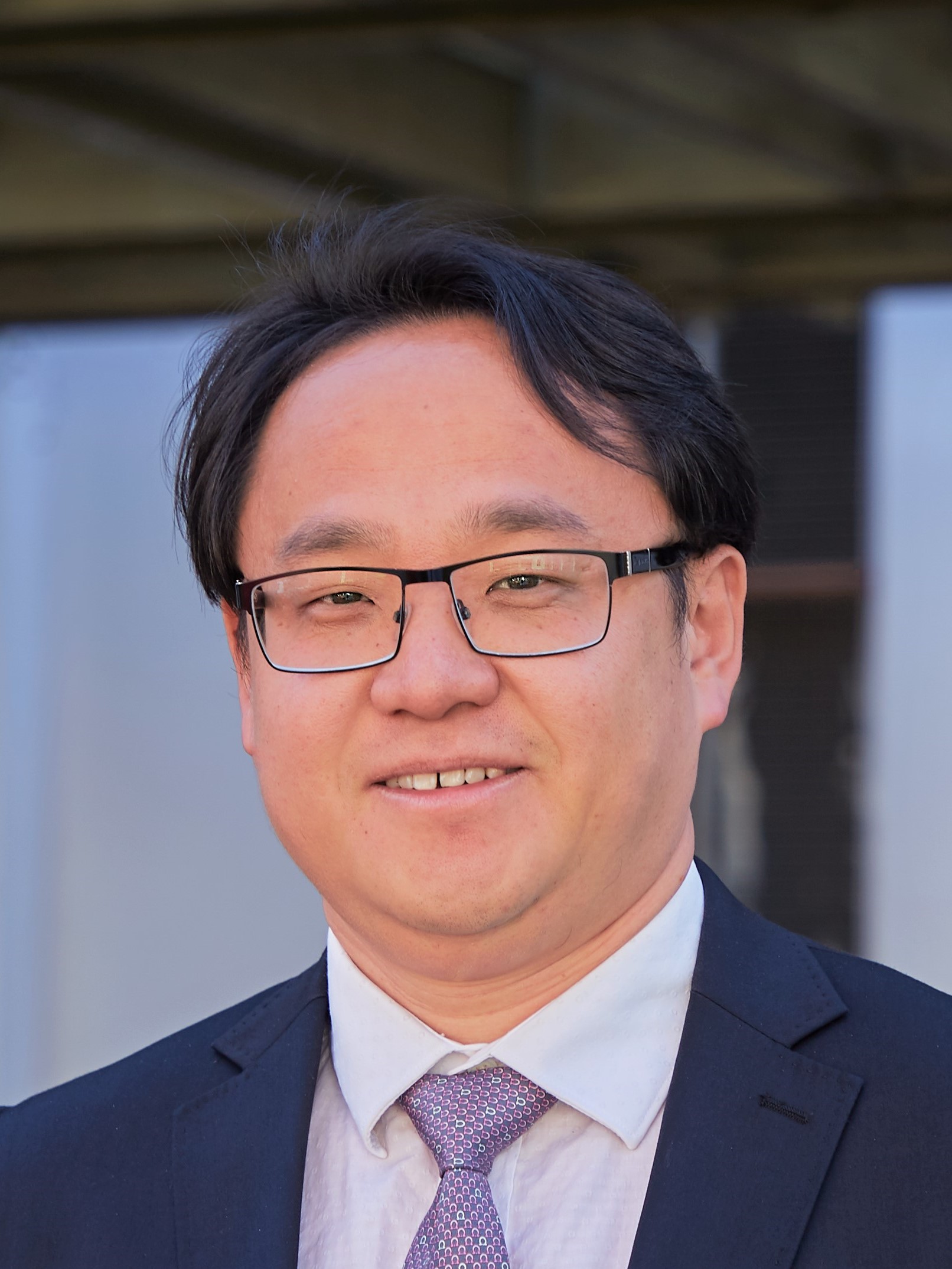}}]
{Nan Yang} (S'09--M'11--SM'18) received the B.S. degree in electronics from China Agricultural University in 2005, and the M.S. and Ph.D. degrees in electronic engineering from the Beijing Institute of Technology in 2007 and 2011, respectively. He has been with the Research School of Electrical, Energy and Materials Engineering at the Australian National University since July 2014, where he currently works as a Senior Lecturer. Prior to this, he was a Postdoctoral Research Fellow at the University of New South Wales (2012--2014) and a Postdoctoral Research Fellow at the Commonwealth Scientific and Industrial Research Organization (2010--2012). He received the IEEE ComSoc Asia-Pacific Outstanding Young Researcher Award in 2014 and the Best Paper Awards from the IEEE GlobeCOM 2016 and the IEEE VTC 2013-Spring. He also received the Top Editor Award from the Transactions on Emerging Telecommunications Technologies, the Exemplary Reviewer Awards from the \textsc{IEEE Transactions on Communications}, \textsc{IEEE Wireless Communications Letters}, and \textsc{IEEE Communications Letters}, and the Top Reviewer Award from the \textsc{IEEE Transactions on Vehicular Technology} from 2012 to 2019. He is currently serving in the Editorial Board of the \textsc{IEEE Transactions on Wireless Communications}, \textsc{ IEEE Transactions on Molecular, Biological, and Multi-Scale Communications}, \textsc{IEEE Transactions on Vehicular Technology}, and \textsc{Transactions on Emerging Telecommunications Technologies}. His research interests include millimeter wave and terahertz communications, ultra-reliable low latency communications, cyber-physical security, massive multi-antenna systems, and molecular communications.
\end{IEEEbiography}
\vspace{-60mm}
\begin{IEEEbiography}[{\includegraphics[width=1in,height=1.25in,clip,keepaspectratio]{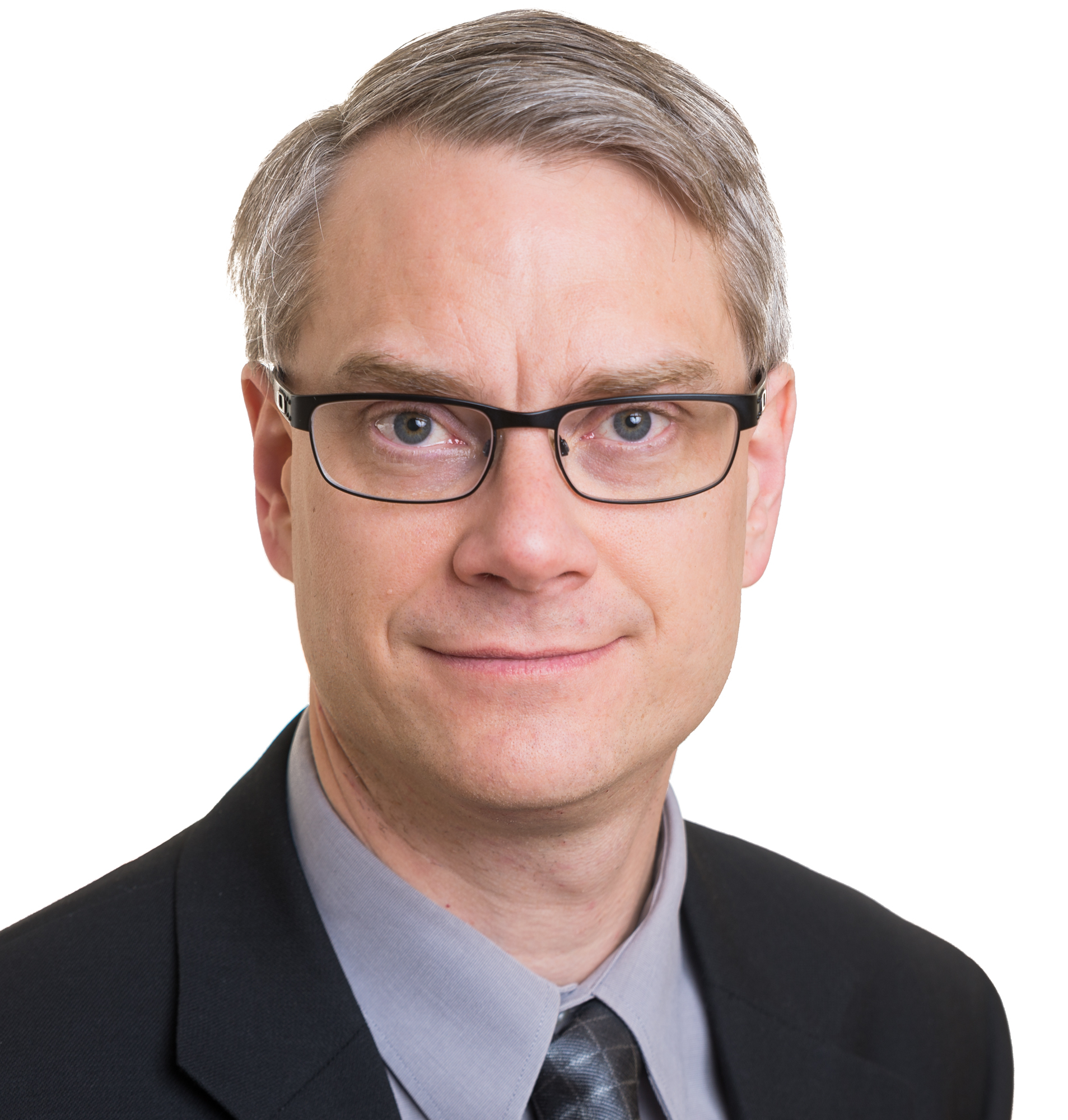}}]
{Erik G. Larsson} (S'99--M'03--SM'10--F'16) received the Ph.D. degree from Uppsala University, Uppsala, Sweden, in 2002.  He is currently Professor of Communication Systems at Link\"oping University (LiU) in Link\"oping, Sweden. He was with the KTH Royal Institute of Technology in Stockholm, Sweden, the George Washington University, USA, the University of Florida, USA, and Ericsson Research, Sweden.  His main professional interests are within the areas of wireless communications and signal processing. He co-authored \emph{Space-Time Block Coding for  Wireless Communications} (Cambridge University Press, 2003) and \emph{Fundamentals of Massive MIMO} (Cambridge University Press, 2016). He is co-inventor of 19 issued U.S. patents.

Currently he is an editorial board member of the \emph{IEEE Signal Processing Magazine}, and a member of the  \emph{IEEE Transactions on Wireless Communications} steering committee. He served as chair of the IEEE Signal Processing Society SPCOM technical committee (2015--2016), chair of  the \emph{IEEE Wireless  Communications Letters} steering committee (2014--2015), technical chair of the Asilomar SSC conference (2015, 2012), technical co-chair of the IEEE Communication Theory Workshop (2019), and member of the  IEEE Signal Processing Society Awards Board (2017--2019). He was Associate Editor for, among others, the \emph{IEEE Transactions on Communications} (2010-2014) and the \emph{IEEE Transactions on Signal Processing} (2006-2010).

He received the IEEE Signal Processing Magazine Best Column Award twice, in 2012 and 2014, the IEEE ComSoc Stephen O. Rice Prize in Communications Theory in 2015, the IEEE ComSoc Leonard G. Abraham Prize in 2017, the IEEE ComSoc Best Tutorial Paper Award in 2018, and the IEEE ComSoc Fred W. Ellersick Prize in 2019.
\end{IEEEbiography}

\end{document}